\documentclass[preprint,superscriptaddress,nofootinbib,preprintnumbers,amsmath,amssymb]{revtex4-1}
\usepackage{bbm}
\usepackage{stmaryrd}
\usepackage[bb=boondox]{mathalfa}
\usepackage{amsfonts}
\usepackage{mathrsfs}
\usepackage{leftidx}
\usepackage{amssymb}
\usepackage{placeins}
\usepackage{relsize}
\usepackage{slashed}
\usepackage{multirow}
\usepackage[dvipsnames]{xcolor}
\usepackage[colorlinks]{hyperref}
\usepackage{epsfig}
\usepackage{graphicx}               % Standard graphics package
\usepackage{url}
\usepackage{float}

\usepackage{adjustbox}
\usepackage{tikz}
\usetikzlibrary{quantikz}

\makeatletter
\newcommand*{\ovA}[1]{%
  \m@th\overline{\mbox{$#1$}\raisebox{2.25mm}{}}%
}
\newcommand*{\ovB}[1]{%
  \m@th\overline{\mbox{$#1$}\raisebox{2.5mm}{}}%
}
\newcommand{\overleftrightsmallarrow}{\mathpalette{\overarrowsmall@\leftrightarrowfill@}}
\newcommand{\overrightsmallarrow}{\mathpalette{\overarrowsmall@\rightarrowfill@}}
\newcommand{\overleftsmallarrow}{\mathpalette{\overarrowsmall@\leftarrowfill@}}
\newcommand{\overarrowsmall@}[3]{%
  \vbox{%
    \ialign{%
      ##\crcr
      #1{\smaller@style{#2}}\crcr
      \noalign{\nointerlineskip}%
      $\m@th\hfil#2#3\hfil$\crcr
    }%
  }%
}
\def\smaller@style#1{%
  \ifx#1\displaystyle\scriptstyle\else
    \ifx#1\textstyle\scriptstyle\else
      \scriptscriptstyle
    \fi
  \fi
}

\makeatother

\def\TFD{{\rm TFD}}
\def\LPS{{\rm \Delta}}

\def\fSWAP{{\rm SWAP} } 
\def\fFT{{\rm FT} } 
\def\Bog{{\rm B}} 

\def\SG{{\rm S}}
\def\HG{{\rm H}}

\def\be{\begin{equation}}
\def\ee{\end{equation}}
\def\bea{\begin{align}}
\def\eea{\end{align}}

\newcommand{\Tr}{ {\rm Tr} }

\begin{document}

\newenvironment{psmallmatrix}
  {\left(\begin{smallmatrix}}
  {\end{smallmatrix}\right)}

%--------------------   Previous format   ----------------------------

\title{Quantum Circuits for Scale Transformations\vspace{0.5cm}}
%\author{Ann E. Nelson}
%\email{aenelson@uw.edu}
%\affiliation{Department of Physics, University of Washington, Seattle, WA}  
\author{Michael Park}
\email{q1park@gmail.com}
%\affiliation{Ronin}  
\affiliation{Department of Physics, University of Washington, Seattle, WA\footnote{work started during appointment}\vspace{1cm}}

\begin{abstract}
\vspace{0.25cm}
In this note we develop quantum circuits for exactly simulating the thermal properties of the quantum XY/Ising chain. These circuits are applicable to the simplest integrable lattice models for which the exact momentum-space scattering-matrix can be constructed in terms of local two-body gates. We discuss a circuit analogy for Wilsonian renormalization group flow and show how it can be used to simulate a global coarse graining transformation. We then construct a circuit that transforms the ground state of these models into the thermo-field double state, and show how it can be used to implement a local scale transformation.
\end{abstract}

\maketitle
%\tableofcontents

%%%%%%%%%% INTRODUCTION %%%%%%%%%%
\section{Introduction}

The distant dream of one day simulating complex quantum systems on quantum computational devices has recently grown more motivating. On one hand the recent advent of noisy-intermediate-scale-quantum (NISQ) \cite{Preskill2018quantumcomputingin} hardware has generated great interest in exploring resource-conscious methods for performing efficient simulations on $\mathcal{O}({\rm few})$ qubits. At the same time, quantum information theoretic approaches to fundamental physics have generated intriguing insights into the quantum description of black holes and the nature of holographic theories. This line of questioning, which began with the black hole information paradox \cite{Hawking:1974sw}, has produced a growing consensus about information conservation (via unitary evolution of quantum systems) as the most robust conservation law in nature. Subsequent studies into the entanglement properties of black holes have led to a revolution in our understanding of the event horizon \cite{Page:1993df,  Susskind:1993if, Hayden:2007cs, Sekino:2008he, Susskind:2014rva, Susskind:2018fmx}, as well new paradoxes about the encoding of degrees of freedom in the black hole interior   \cite{Almheiri:2012rt, Susskind:2012rm, Maldacena:2013xja, Harlow:2013tf}.

There is a rich literature of recipes for mapping the mathematical space of interesting physical observables to the computational space of qubits \cite{Kogut:1974ag,Kitaev:2000ak,Jordan:2011ci}. These mappings are typically simple for fermionic systems due to the Pauli exclusion principle, however progress has also been made on the simulation of bosons and their effective interactions \cite{Macridin:2018gdw,Macridin:2018oli,Klco:2018kyo,Klco:2018zqz, Kaplan:2018vnj}. In general the task of finding unitary operators for implementing the exact time evolution of a given quantum system is prohibitively difficult and most practical simulations rely on the Suzuki-Trotter approximation \cite{1990PhLA..146..319S,1991JMP....32..400S}. However exact constructions \cite{Verstraete:2008qpa,Schmoll:2016kbb,Jiang:2017pyp} are known for some interesting integrable models such as the XY chain \cite{Lieb:1961fr,Pfeuty:1970pf} considered here and the Kitaev honeycomb \cite{Kitaev:2006lla}, thus enabling an exact simulation of dynamics in these theories. In this note we build on these algorithms by constructing circuits for performing global and local scale transformations, and simulating the dynamical evolution of thermal states. 

Many of the operations considered here are analogous to those that have been studied in the context of tensor networks such as the multi-scale-renormalization-ansatz (MERA) \cite{Vidal:2003lvx,Vidal:2007hda,Evenbly:2007hxg}. The MERA is a computationally efficient variational ansatz for simulating correlations in strongly interacting many-body systems. The two-body local structure of the MERA naturally defines a notion of causality in which time evolution is represented by transformations of scale. A holographic interpretation of the MERA at criticality was shown to bear a striking resemblance to the entanglement structure of higher dimensional theories with a deSitter metric \cite{Swingle:2009bg,Beny:2011vh,SinaiKunkolienkar:2016lgg}. Later the MERA was reinterpreted in the context of the duality between conformal field theories (CFT's) and ``kinematic space" in which the conditional mutual information between pairs of qubits in the CFT correspond to the volume element of the dual space \cite{Czech:2015qta,Czech:2015kbp}. These authors showed in \cite{Czech:2015xna} that a discrete logarithmic mapping on the MERA produces the thermal state, thus recovering a discrete version of the local scale invariance enjoyed by conformally invariant theories in the $1+1$ dimensional continuum. These intruiging connections have even led to studies on the cosmological implications of a MERA-type universe \cite{Bao:2017qmt,Bao:2017iye}.

This note is organized as follows. In Section \ref{sec:XYchain} we describe the quantum XY chain and review the procedure for diagonalizing its Hamiltonian. In Section \ref{sec:XYcircuit} we describe the unitary circuit that performs this diagonalization, and show how these circuits can be used to perform a coarse graining procedure on the simulated space of states by taking a partial trace over microscopic degrees of freedom. In Section \ref{sec:thermal} we describe an algorithm for constructing the thermo-field double ($\TFD$) state. The $\TFD$ is a state of great interest for holographic theories such as AdS/CFT, where it is known to be the boundary state which is dual to a two-sided black hole in the bulk. We study the entanglement properties of the $\TFD$ state at the quantum critical point and find evidence that it corresponds to a logarithmic rescaling of position-space coordinates. In Section \ref{sec:summary} we summarize our results.

\section{The Quantum XY Chain}
\label{sec:XYchain}

The quantum XY chain is one of the simplest exactly-solvable many-body systems, consisting of a nearest-neighbor interaction with strength $J$ and a transverse magnetic field interaction with strength $\Gamma$. The dynamics of a (anti)periodic chain of $n$ qubits is governed by the Hamiltonian in Eq \ref{eq:XYpauli}. Here we use the standard notation for quantum gates in which $X_i$, $Y_i$, $Z_i$ refer to the corresponding $2 \times 2$ Pauli matrices acting on the $i^{\rm th}$ qubit. The term $H_\partial$ corresponds to a boundary term, which we assume to be periodic thus restricting to $n$ even. 

\be
H_{XY} = \sum_{j = 1}^{n} \Bigg[ J \left( \frac{1 + \gamma}{2} \, X_j X_{j+1} + \frac{1 - \gamma}{2} \, Y_j Y_{j+1} \right) + \Gamma \, Z_j - \frac{n}{2} \, I \Bigg] + H_{\partial}
\label{eq:XYpauli}
\ee
The physical properties of this Hamiltonian can be made more manifest by transforming it from position space to the energy eigenbasis. The Hamiltonian is diagonal in this basis and the eigenstates evolve as non-interacting particles with the momentum-dependent single-particle frequencies $w_p$ given in Eq \ref{eq:1parw}. In terms of the energy eigenstates $\eta_p$ the Hamiltonian takes the form in Eq \ref{eq:XYe}

\be
\widetilde{H}_{XY} = \displaystyle\sum_p w_p \eta^\dagger_p \eta_p
\label{eq:XYe}
\ee
\be
w_p(\lambda, \gamma) = \sqrt{ (1 - \lambda \cos 2 \pi p )^2 + (\lambda \, \gamma \sin 2 \pi p)^2 }
\label{eq:1parw}
\ee
Here we have rescaled the Hamiltonian by the transverse field setting $\Gamma = 1$ and expressed the frequencies in terms of the ratio $\lambda = J / \Gamma$.

The diagonalization procedure occurs in three steps. The first step is a simplification via Jordan-Wigner transform to the fermionic representation. In terms of the fermionic annihilation operators defined in Eq \ref{eq:fermc}, the Hamiltonian takes the form in Eq \ref{eq:XYf}

\be
c_j =  \bigg( \displaystyle\prod_{k<j} Z_k \bigg) \frac{X_j + i Y_j}{2} 
\label{eq:fermc}
\ee
\be
H_{XY} = \sum_{j = 1}^{n} \frac{\lambda}{2} \Big[ \left( c_j^\dagger c_{j+1} + c_{j+1}^\dagger c_j \right) + \gamma \left( c_j c_{j+1} + c_{j+1}^\dagger c_j^\dagger \right) \Big] + c_j^\dagger c_j
\label{eq:XYf}
\ee
The second step is to perform discrete Fourier transform trading the position space fermions $c_j$ for momentum space fermions $\tilde{c}_p$. The Fourier transform is manifestly unitary and thus defines a matrix $U_{\rm FT}$ acting on the the $c_j$ as shown in Eq \ref{eq:UFT}.

\be
U_{\rm FT}^\dagger \, c_j \, U_{\rm FT} = \tilde{c}_p \qquad \text{where} \qquad \tilde{c}_p = \frac{1}{\sqrt{n}} \displaystyle\sum_{j = 1}^n e^{2 \pi i p j} c_j
\label{eq:UFT}
\ee
$$
p \equiv \frac{k}{n} \qquad \qquad k = \left( +1, -1, +2, -2, \ldots, n-1, -n+1, n, 0 \right)
$$
The final step is a Bogoliubov transformation that trades the momentum space fermions $\tilde{c}_p$ for the energy eigenstates $\eta_{p}$. In these simplest integrable models, the energy eigenstates are simply entangled pairs of left/right-moving modes. The unitary matrix implementing the Bogoliubov transformation can thus be expressed as a tensor product of two-qubit gates $\Bog_p$ as shown in Eq \ref{eq:UBog}. Note that the gate $\Bog_p$ shown in Eq \ref{eq:Bog2} (dis)entangles pairs of opposite-momentum modes and is defined with respect to the absolute value $|p|$. An explicit construction for $\Bog_p$ in terms of standard gates is given in Appendix \ref{app:circuits}.

\be
U_{\rm Bog}^\dagger \, \tilde{c}_p \, U_{\rm Bog} = \eta_{p} \qquad \text{where} \qquad U_{\rm Bog} = \bigotimes_{(p, -p)} {\Bog}_p
\label{eq:UBog}
\ee

\be
 \qquad \qquad {\Bog}_p = \begin{pmatrix}
\cos \frac{\theta_p}{2} & 0 & 0 & i \sin \frac{\theta_p}{2} \\
0 & ~1~ & ~0~ & 0 \\
0 & 0 & 1 & 0 \\
i \sin \frac{\theta_p}{2} & 0 & 0 & \cos \frac{\theta_p}{2} 
\end{pmatrix} \qquad \qquad \theta_p(\lambda, \gamma) = \cos^{-1}\left( \frac{-1 + \lambda \cos 2 \pi p }{w_p(\lambda, \gamma)} \right) \quad %\pi - 
\label{eq:Bog2}
\ee
Thus the composition of the Fourier transform with the Bogoliubov transformation defines a unitary matrix $U_{\rm Dis}$, which can be used to move states between the position and energy eigenbases as shown in Eq \ref{eq:HDis}.

\be
\widetilde{H}_{XY} = U_{\rm Dis}^\dagger H_{XY} U_{\rm Dis} \qquad \text{where} \qquad U_{\rm Dis} = U_{\rm FT} U_{\rm Bog}
\label{eq:HDis}
\ee

\section{A Quantum XY Circuit}
\label{sec:XYcircuit}

The construction of an explicit circuit for diagonalizing the XY Hamiltonian in terms of two-body local gates was given in \cite{Verstraete:2008qpa}, and later developed in \cite{Schmoll:2016kbb,Jiang:2017pyp}. Their circuit follows the prescription of the analytic diagonalization procedure, with a Fourier transform into momentum space followed by a Bogoliubov transformation that (dis)entangles left/right-moving modes. The abstracted circuit structure is shown in Fig \ref{diag:dis}. Here we assume that the Fourier transform gate $\fFT$ is defined modulo products of two-body $\fSWAP$ operations, which correspond to a reordering of the momentum modes $\ket{p}$ in the space of qubits. An explicit construction of $\fFT$  in terms of two-body local gates is given in Appendix \ref{app:circuits}. Here we emphasize that the simplicity of this circuit relies on the fact that transforming from the momentum to energy eigenbasis involves a simple $SU(2)$ rotation between momentum modes of opposite sign, which is implemented by the Bogoliubov transformation. These Bogoliubov transformations can thus be interpreted as the simplest non-trivial example of a momentum-space scattering matrix. In principle more complex theories could be simulated by replacing the $\Bog_p$ gates with a sequence of gates that implement scattering effects between momentum modes of different magnitudes. For example, any integrable model that can be analytically solved using the Bethe ansatz should also have a representation in terms of two-body local gates.

\begin{center}
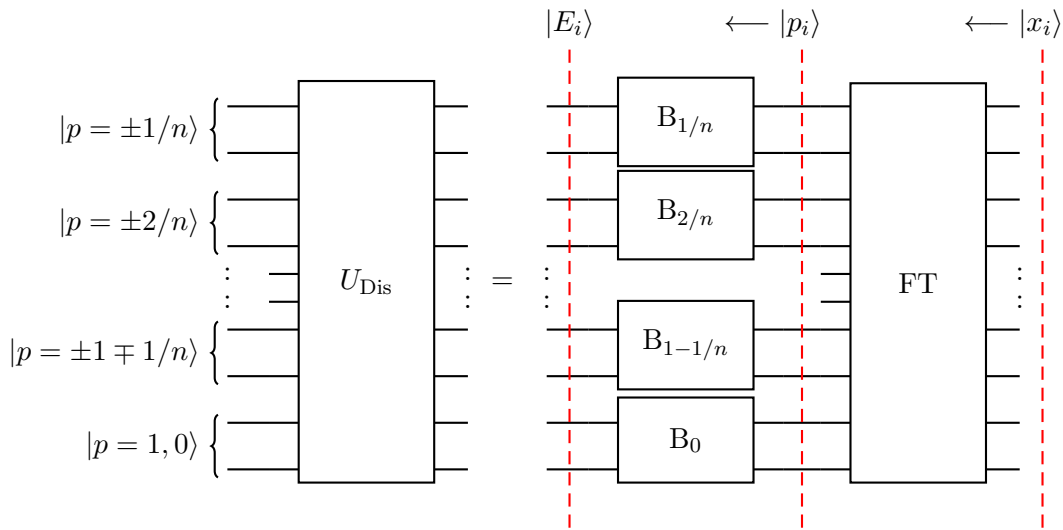
\begin{figure}[ht]
\begin{tikzcd}[row sep={0.62cm,between origins}]
& & & \\
\lstick[wires=2]{$\ket{p=\pm 1/n}$} & \qw & \gate[wires=10]{U_{\rm Dis} } & \qw \\ 
& \qw & & \qw \\ 
\lstick[wires=2]{$\ket{p=\pm 2/n}$} & \qw & & \qw \\
& \qw & & \qw \\[-0.25cm]
: & & & : \\[-0.25cm]
: & & & : \\[-0.25cm]
\lstick[wires=2]{$\ket{p=\pm 1 \mp 1/n}$} & \qw & & \qw \\
& \qw & & \qw \\
\lstick[wires=2]{$\ket{p=1, 0}$} & \qw & & \qw \\
& \qw & \hphantom{wide label} & \qw \\
& & &
\end{tikzcd}
=
\begin{tikzcd}[row sep={0.62cm,between origins}]
\slice{$\ket{E_i}$} & & & \slice{$\longleftarrow \ket{p_i} \qquad$} & & & \slice{$\longleftarrow \ket{x_i} \qquad$} & \\
& \qw & \gate[wires=2]{\Bog_{1/n}} & \qw & \qw & \gate[wires=10]{\fFT} & \qw & \\ 
& \qw & \hphantom{wide label} & \qw & \qw & & \qw & \\ 
& \qw & \gate[wires=2]{\Bog_{2/n}}   & \qw & \qw & & \qw & \\
& \qw & \hphantom{wide label} & \qw & \qw & & \qw & \\[-0.25cm]
: & & & & & & : & \\[-0.25cm]
: & & & & & & : & \\[-0.25cm]
& \qw & \gate[wires=2]{\Bog_{1-1/n}}   & \qw & \qw & & \qw & \\
& \qw & \hphantom{wide label} & \qw & \qw & & \qw & \\
& \qw & \gate[wires=2]{\Bog_{0}}   & \qw & \qw & & \qw & \\
& \qw & \hphantom{wide label} & \qw & \qw & \hphantom{wide label} & \qw & \\
& & & & & & &
\end{tikzcd}
\caption{A quantum circuit that sequentially transforms position eigenstates $\ket{x_i}$ to momentum eigenstates $\ket{p_i}$ to energy eigenstates $\ket{E_i}$. A Fourier transform ${\fFT}$ followed by ${\rm SWAP}$ operations can be used arrange modes with $\pm p$ into nearest-neighbor pairs. Two-qubit Bogoliubov transformations are then used to (dis)entangle the left/right moving modes.}
\label{diag:dis}
\end{figure}
\end{center}
\vspace{-0.5cm}
The unitary disentangling operation illustrated in Fig \ref{diag:dis} can be used to transform energy eigenstates represented on a trivial product state into a highly entangled sum of position space modes, and vice-versa via $\ket{x_i} = U_{\rm Dis} \ket{E_i}$. The time evolution of energy eigenstates under the diagonalized Hamiltonian can be simulated by applying single-qubit unitary rotations as shown in Fig \ref{diag:time}.
\begin{center}
\begin{figure}[ht]
\begin{tikzcd}[row sep={0.85cm,between origins}]
& & & \\ 
& \qw & \gate[wires=6]{\quad U(t) \quad} & \qw \\
& \qw & \qw & \qw \\[-0.45cm]
: & & & : \\[-0.45cm]
: & & & : \\[-0.45cm]
& \qw & \qw & \qw \\
& \qw & \qw & \qw \\
& & &
\end{tikzcd}
\quad =
\begin{tikzcd}[row sep={0.85cm,between origins}]
\slice{$\ket{E_i (t)}$} & & & \slice{$\longleftarrow \ket{E_i (0)} \qquad$} & \\ 
& \qw & \gate{\quad ~e^{- \frac{i}{2} w_{1/n} t }~ \quad} & \qw & \\ 
& \qw & \gate{\quad e^{- \frac{i}{2} w_{-1/n} t } \quad} & \qw & \\[-0.45cm]
: & & & : & \\[-0.45cm]
: & & & : & \\[-0.45cm]
& \qw & \gate{\quad ~~e^{- \frac{i}{2} w_1 t }~~ \quad} & \qw & \\
& \qw & \gate{\quad ~~e^{- \frac{i}{2} w_0 t }~~ \quad} & \qw & \\
& & & &
\end{tikzcd}
\caption{A quantum circuit for implementing the time-evolution of energy eigenstates. Since the energy eigenstates in the XY model behave as non-interacting particles, the unitary time-evolution operator can be represented as a product of single-qubit phase rotations.}
\label{diag:time}
\end{figure}
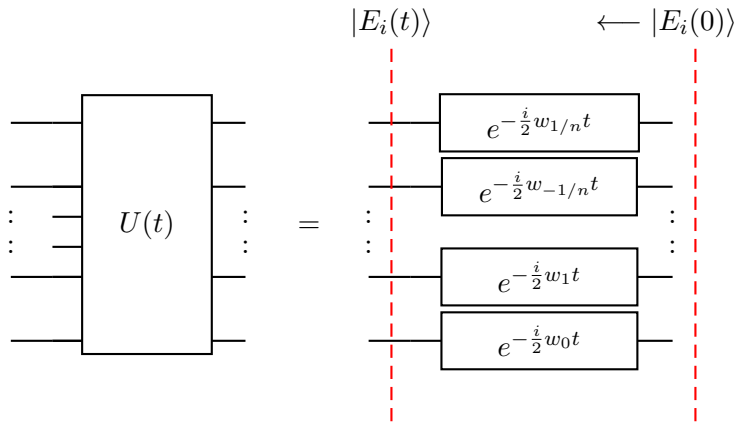
\end{center}
\vspace{-0.5cm}

As a simple demonstration we can start with the trivial product state representing a definite position, then transform it to the energy basis and apply the time evolution operator for some time $t$, then transform back to position space. This is equivalent to the sequence of operations in Eq \ref{eq:evolvex}. In Fig \ref{diag:RG}  (top) we show the time evolution of the expectation value $\langle Z_i \rangle$ in the critical Ising chain for initial states corresponding to various eigenstates of $Z_i$.

\be
\ket{x_i (t)} = U_{\rm Dis} \, e^{-\frac{i}{2} \widetilde{H}_{XY} \Delta t} \, U_{\rm Dis}^\dagger \ket{x_i (0)}
\label{eq:evolvex}
\ee

\begin{center}
\begin{figure}[ht]
  \begin{center}
    \includegraphics[scale=0.425]{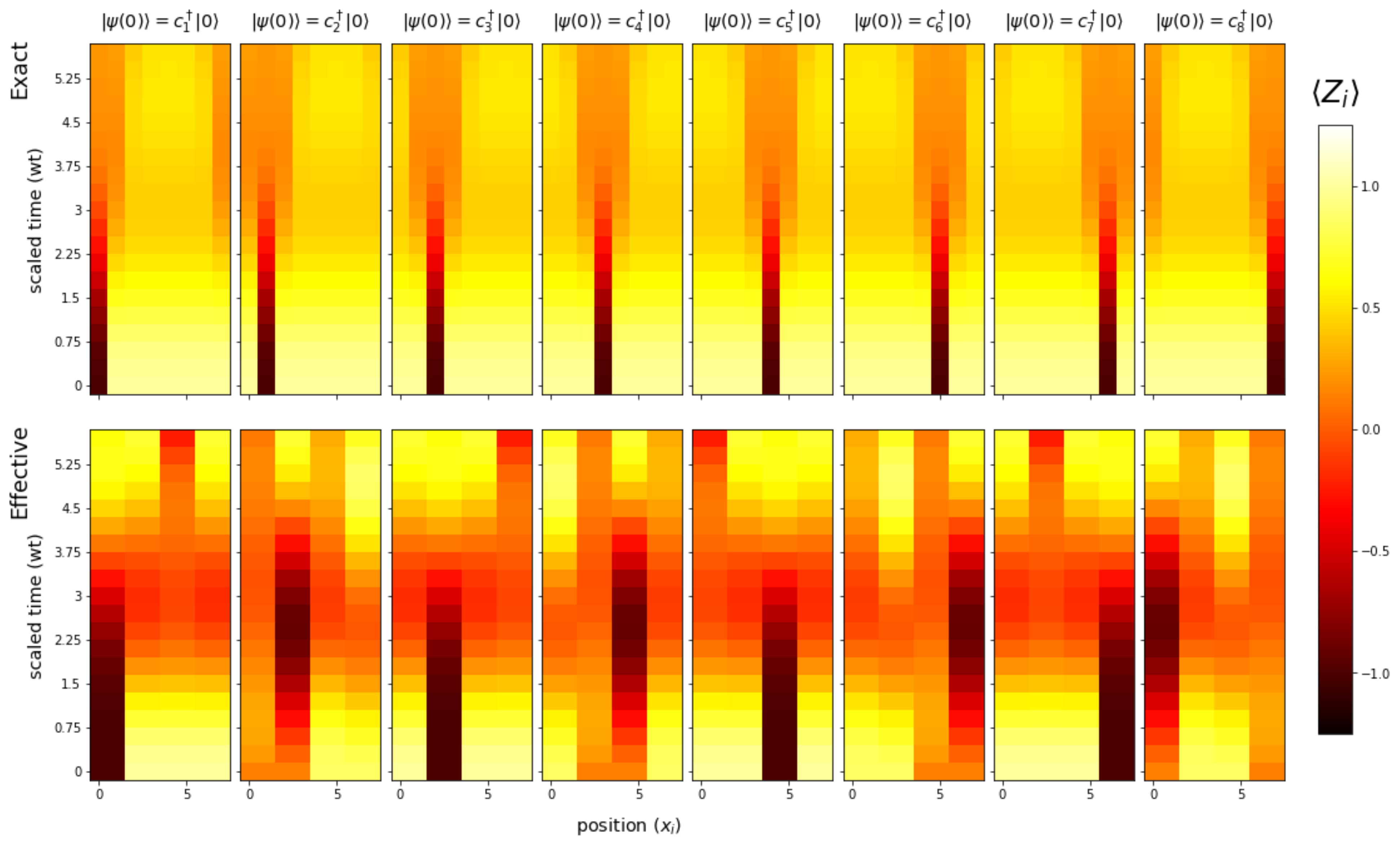}
  \end{center}
  \caption{The spacetime evolution of the expectation value $\langle Z_i \rangle$, starting with single-particle initial states of definite position. The top row shows the evolution of position states for a fine-grained simulation on $n=8$ qubits. The bottom row shows the evolution for a coarse-grained effective theory on $n=4$ qubits, obtained by taking a partial trace over high-momentum modes.}
  \label{fig:course_expZ}
\end{figure}
\end{center}
\vspace{-0.5cm}
The simplicity of the XY model allows for great control in manipulating its degrees of freedom. This allows us to define coarse graining procedures that are analogous to those familiar from quantum field theory. For example position-space renormalization with a hard momentum cutoff can be performed in the obvious way by taking a partial trace over momentum modes greater than some ultraviolet cutoff $\Lambda$. In order to keep the construction restricted to two-body local gates, the high and low momentum modes should be physically separated using a product of two-local SWAP operations following the Fourier transform as shown in Fig \ref{diag:RG}. An inverse Fourier transform on the first $p$ qubits completes the transformation to the position-space state of a coarse grained theory on the first $p$ qubits. An explicit construction of the SWAP operations that perform this reordering is given in Appendix \ref{app:circuits}. The time evolution of position eigenstates in the coarse-grained effective theory is shown in Fig \ref{fig:course_expZ} (bottom). The coarse grained density matrix $\rho_{\rm IR}$ is thus related to the fine-grained density matrix $\rho_{\rm UV}$ via Eq \ref{eq:IRUV}. Given two pure state density matrices $\rho_A$ and $\rho_B$, the invariance of the inner product of states $\Tr (\rho_A \rho_B)$ under this procedure guarantees that it describes an isometric transformation.

\be
\rho_{\rm IR} = \Tr_{p>\Lambda} \left( U_{\rm RG} \, \rho_{\rm UV} \, U_{\rm RG}^\dagger \right)
\label{eq:IRUV}
\ee

\begin{center}
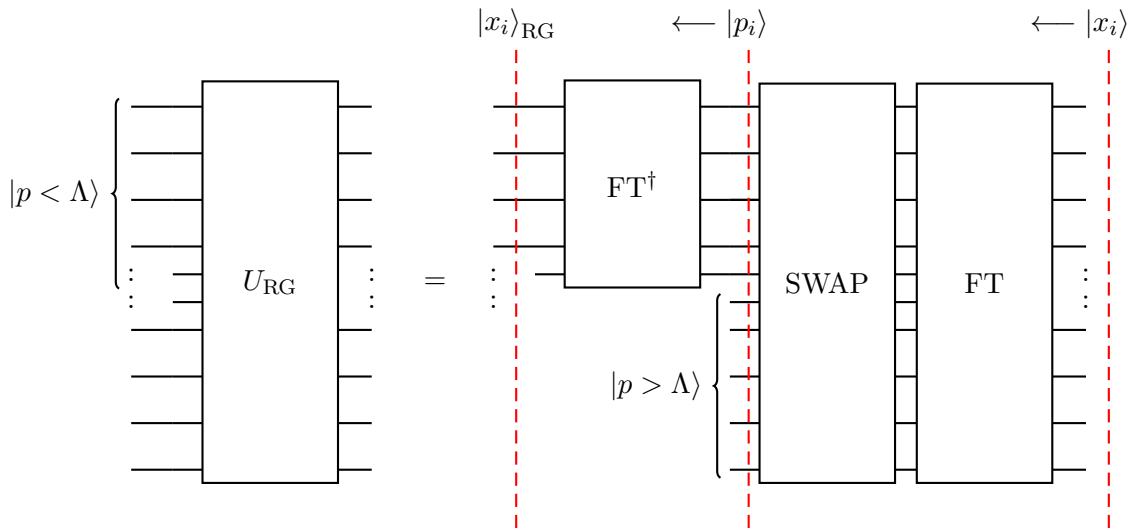
\begin{figure}[ht]
\begin{tikzcd}[row sep={0.62cm,between origins}]
& & & \\
\lstick[wires=5]{$\ket{p<\Lambda}$} & \qw & \gate[wires=10]{U_{\rm RG} } & \qw \\ 
& \qw & & \qw \\ 
& \qw & & \qw \\
& \qw & & \qw \\[-0.25cm]
: & & & : \\[-0.25cm]
: & & & : \\[-0.25cm]
& \qw & & \qw \\
& \qw & & \qw \\
& \qw & & \qw \\
& \qw & \hphantom{wide label} & \qw \\
& & &
\end{tikzcd}
\quad=
\begin{tikzcd}[row sep={0.62cm,between origins}]
\slice{$\ket{x_i}_{
\rm RG}$} & & & \slice{$\longleftarrow \ket{p_i} \qquad$} & & & \slice{$\longleftarrow \ket{x_i} \qquad$} & \\
& \qw & \gate[wires=5]{\fFT^\dagger} & \qw & \gate[wires=10]{\fSWAP} & \gate[wires=10]{\fFT} & \qw & \\ 
& \qw & & \qw & & & \qw & \\ 
& \qw & & \qw & \qw & & \qw & \\
& \qw & & \qw & & & \qw & \\[-0.25cm]
: & & \hphantom{wide label} & \qw & & & : & \\[-0.25cm]
: & & & \lstick[wires=5]{\ket{p>\Lambda}} & & & : & \\[-0.25cm]
& & & & & & \qw & \\
& & & & & & \qw & \\
& & & & & & \qw & \\
& & & & \hphantom{wide label} & \hphantom{wide label} & \qw & \\
& & & & & & &
\end{tikzcd}
\caption{A unitary gate that localizes the high-momentum modes in position-space. The SWAP operation which rearranges momentum eigenstates is needed only to keep the construction restricted to two-body local gates. In practice one may just keep track of the high momentum modes and perform non-local operations over them.}
\label{diag:RG}
\end{figure}
\end{center}
\vspace{-0.5cm}

\section{Entangling Thermal States}
\label{sec:thermal}

\begin{center}
\begin{figure}[ht]
  \begin{center}
    \includegraphics[scale=0.85]{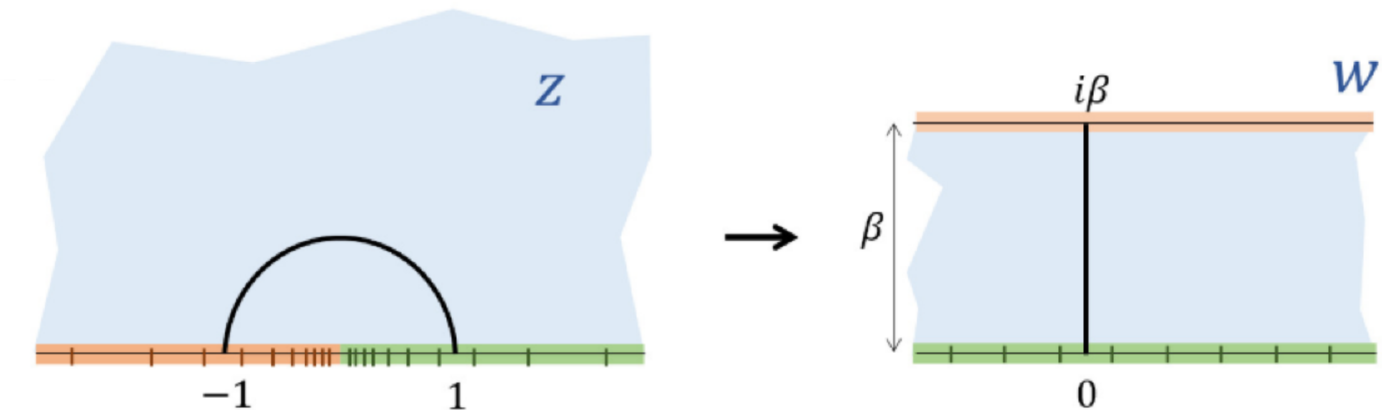}
  \end{center}
  \caption{Left: The Euclidean path integral on the upper half plane (blue) prepares the ground state on an infinite line (red/green). A Weyl rescaling maps the upper half plane to an infinite strip with a width determined by the inverse temperature $\beta$. Right: The Euclidean path integral on this strip prepares two copies the thermal state on the infinite line.}
  \label{fig:TFD_CFT}
\end{figure}
\end{center}

In this section we construct a circuit for transforming a representation of the ground state into the thermo-field double (\TFD) \cite{Takahashi:1996zn}. The $\TFD$ state is a highly entangled pure state that is bifurcated into two subsystems $L$ and $R$ in such a way that it appears exactly thermal to observers with access to $L$ ($R$) but not $R$ ($L$). It can thus be defined as the purification of the thermal state and expressed in the energy eigenbasis as in Eq \ref{eq:TFD}. By definition the partial trace of the $\TFD$ state over $L$ ($R$) gives the mixed thermal state on $R$ ($L$), as in Eq \ref{eq:TMS}

\be
\ket{\TFD} \equiv \displaystyle\sum_{i = 1}^{2^{n-1} } \sqrt{\frac{e^{- \beta E_i} }{Z}} \, \ket{E_i}_L \ket{E_i}_R \qquad \qquad \beta \equiv \frac{1}{T}
\label{eq:TFD}
\ee

\be
\Tr_R \left( \ket{\TFD} \bra{\TFD} \right) = \displaystyle\sum_{i = 1}^{2^{n-1} } \frac{e^{- \beta E_i} }{Z} \, | E_i \rangle_L \langle E_i |
\label{eq:TMS}
\ee
In $d=1+1$ conformal field theories (CFT's), the $\TFD$ state is related to the ground state by a local conformal map called a Weyl rescaling. The map takes the ground state defined on the punctured real line $\mathbb{R} / \{0\}$ with coordinate $z$, to an entangled state on $\mathbb{R} \times \mathbb{R}$ with coordinate $w$ as defined in Eq \ref{eq:Weyl}. Thus an exponentially spaced discretization of the positive real axis on $z$ is mapped to a regular discretization of the real axis on $w$. 
This procedure is illustrated in Fig \ref{fig:TFD_CFT} which was taken from the original source \cite{Czech:2015xna}.

\be
z \rightarrow w = (\beta / \pi) \log z
\label{eq:Weyl}
\ee
%\vspace{-0.5cm}

The $\TFD$ at inverse temperature $\beta$ can be prepared by first performing a unitary transformation $U_{\LPS} (\beta)$ from the ground state to the ``Laplacian state", which we call $\ket{\LPS}$. This state is simply the Laplace transform of the energy eigenstates, and can be expressed as a product state involving creation operators acting on the vacuum as shown in Eq \ref{eq:LPS}

\begin{align}
\ket{\LPS} &\equiv \displaystyle\sum_{i = 1}^{2^n} \sqrt{\frac{e^{- \beta E_i} }{Z}} \, \ket{E_i} \\
&= \displaystyle\prod_{p = 1/n}^1 \frac{e^{-\frac{\beta w_p}{2}} I + e^{+\frac{\beta w_p}{2}} \tilde{c}_p^\dagger}{\sqrt{e^{- \beta w_p} + e^{+ \beta w_p} } } \ket{E_0}
\label{eq:LPS}
\end{align}
The unitary operator that creates the $\ket{\LPS (\beta)}$ state can thus be expressed as a tensor product of two-qubit gates $L(\beta, w_p)$ as defined in Eq \ref{eq:ULP}. Note that in the limit where $w_p \rightarrow 0$, the $L(\beta, w_p)$ gate reduces to the Hadamard gate. The circuit representation of the unitary transformation that creates the Laplacian state is shown in Fig \ref{diag:LP}. Starting from the ground state $\ket{E_0}$, the $\TFD$ state can be prepared simply by applying the Laplace transform $U_{\LPS} (\beta)$ on half the qubits followed by a sequence of CNOT gates as shown in Fig \ref{diag:TFD}.
\be
L (\beta, w_p) = \frac{1}{\sqrt{e^{- \beta w_p} + e^{+ \beta w_p} } } \left( e^{-\frac{\beta w_p}{2}} X + e^{+\frac{\beta w_p}{2}} Z \right)
\ee
\be
U_{\LPS} (\beta) = \bigotimes_{p = 1/n}^1 L (\beta, w_p)
\label{eq:ULP}
\ee
\begin{center}
\begin{figure}[ht]
\begin{tikzcd}[row sep={0.85cm,between origins}]
& & & \\ 
& \qw & \gate[wires=6]{\quad U_{\LPS} (\beta) \quad} & \qw \\ 
& \qw & & \qw \\[-0.45cm]
: & & & : \\[-0.45cm]
: & & & : \\[-0.45cm]
& \qw & & \qw \\
& \qw & & \qw \\
& & & &
\end{tikzcd}
\quad =
\begin{tikzcd}[row sep={0.85cm,between origins}]
\slice{$\ket{\LPS}$} & & & \slice{$\longleftarrow \ket{E_0} \qquad$} & \\ 
& \qw & \gate{\quad ~L(\beta, w_{1/n})~ \quad} & \qw & \\ 
& \qw & \gate{\quad L(\beta, w_{-1/n}) \quad} & \qw & \\[-0.45cm]
: & & & : & \\[-0.45cm]
: & & & : & \\[-0.45cm]
& \qw & \gate{\quad ~~L(\beta, w_1)~~ \quad} & \qw & \\
& \qw & \gate{\quad ~~L(\beta, w_0)~~ \quad} & \qw & \\
& & & &
\end{tikzcd}
\caption{Quantum circuit for preparing the Laplacian state. The Laplacian state is a product state and can hence be transformed from the ground state by a product of single-qubit $L(\beta, w_p)$ rotations.}
\label{diag:LP}
\end{figure}
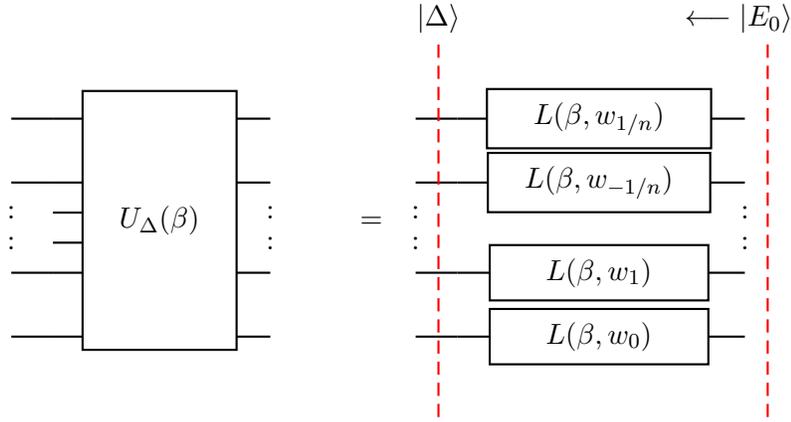
\end{center}
\vspace{-0.5cm}

\begin{center}
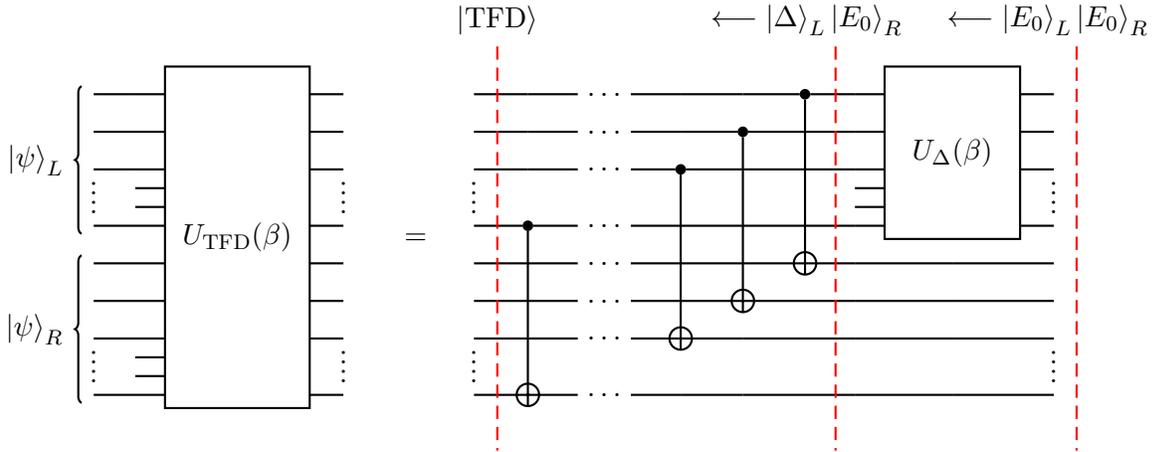
\begin{figure}[ht]
\begin{tikzcd}[row sep={0.5cm,between origins}]
& & & \\ 
\lstick[wires=6]{$\ket{\psi}_L$} & \qw & \gate[wires=12]{~U_{\TFD} (\beta)~} & \qw \\ 
& \qw & & \qw \\
& \qw & & \qw \\[-0.25cm]
: & & & : \\[-0.25cm]
: & & & : \\[-0.25cm]
& \qw & & \qw \\
\lstick[wires=6]{$\ket{\psi}_R$} & \qw & & \qw  \\
& \qw & & \qw \\
& \qw & & \qw \\[-0.25cm]
: & & & : \\[-0.25cm]
: & & & : \\[-0.25cm]
& \qw & & \qw \\
& & & &
\end{tikzcd}
=
\begin{tikzcd}[row sep={0.5cm,between origins}]
\slice{$\ket{\TFD}$} & & & & & \slice{$\longleftarrow \ket{\LPS}_L \ket{E_0}_R \qquad$} & & & \slice{$\longleftarrow \ket{E_0}_L \ket{E_0}_R \qquad$} &  \\
& \qw & \qw\ \ldots\ & \qw & \qw & \ctrl{6} & \qw & \gate[wires=6]{U_{\LPS} (\beta)} & \qw & \\
& \qw & \qw\ \ldots\ & \qw & \ctrl{6} & \qw & \qw & & \qw & \\ 
& \qw & \qw\ \ldots\ & \ctrl{6} & \qw & \qw & \qw & & \qw & \\[-0.25cm]
: & & & & & & & & : & \\[-0.25cm]
: & & & & & & & & : & \\[-0.25cm]
& \ctrl{6} & \qw\ \ldots\ & \qw & \qw & \qw & \qw & \hphantom{wide label} & \qw & \\
& \qw & \qw\ \ldots\ & \qw & \qw & \targ{} & \qw & \qw & \qw & \\ 
& \qw & \qw\ \ldots\ & \qw & \targ{} & \qw & \qw & \qw & \qw & \\ 
& \qw & \qw\ \ldots\ & \targ{} & \qw & \qw & \qw & \qw & \qw & \\[-0.25cm]
: & & & & & & & & : & \\[-0.25cm]
: & & & & & & & & : & \\[-0.25cm]
& \targ{} & \qw\ \ldots\ & \qw & \qw & \qw & \qw & \qw & \qw & \\
& & & & & & & & & &
\end{tikzcd}
\caption{Quantum circuit for preparing the thermo-field double state. The circuit splits into two parts. The first takes the ground state to a product of the Laplacian and ground states. The second uses a sequence of CNOT gates to complete the transformation.}
\label{diag:TFD}
\end{figure}
\end{center}
\vspace{-0.5cm}

\begin{center}
\begin{figure}[ht]
  \begin{center}
    \includegraphics[scale=0.5]{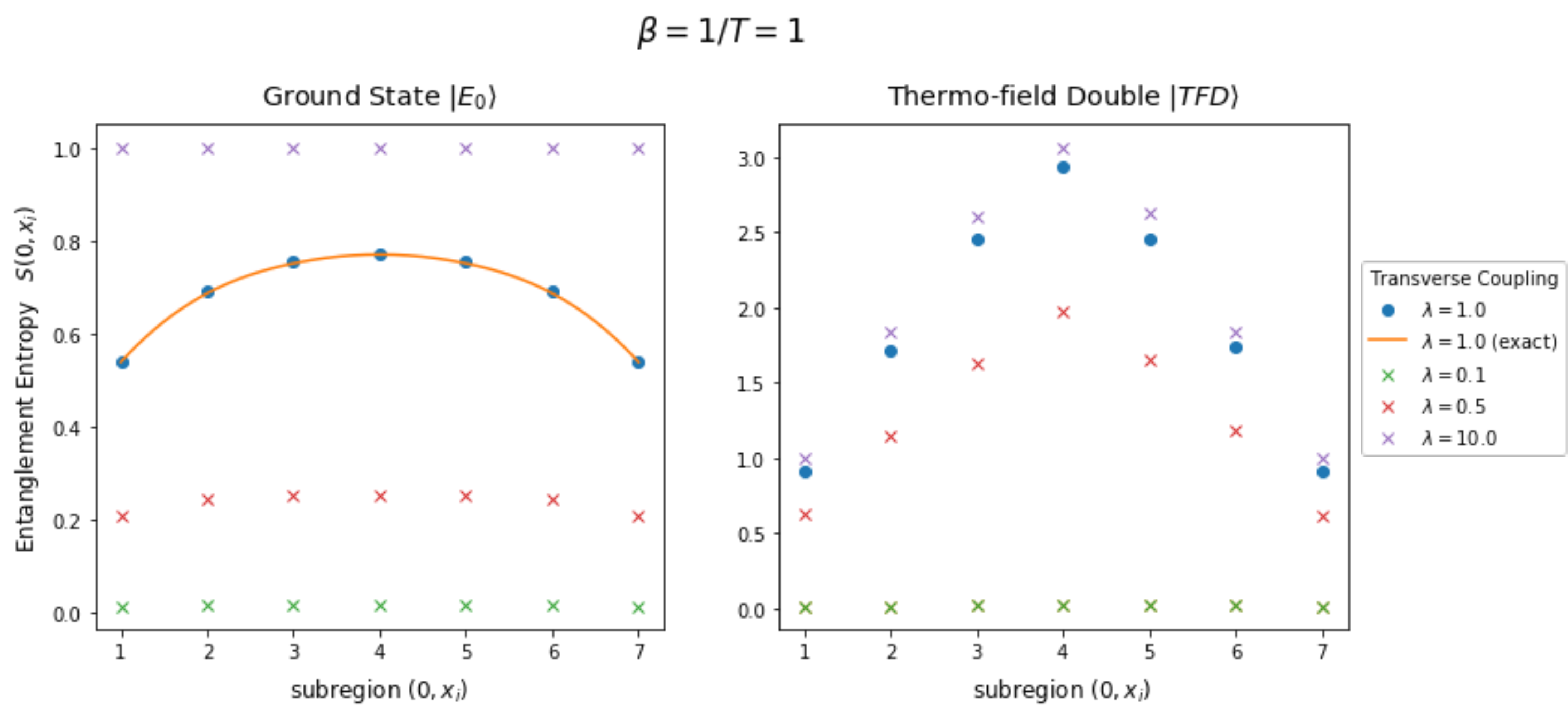}
  \end{center}
  \caption{Entanglement entropy $S(\ell)$ in the ground state versus the thermo-field double. At the quantum critical point $\lambda=1$, the logarithmic scaling of $S(\ell)$ in the vacuum is mapped to a linear scaling in the thermo-field double state.}
  \label{fig:TFD_T1}
\end{figure}
\end{center}

\begin{center}
\begin{figure}[ht]
  \begin{center}
    \includegraphics[scale=0.6]{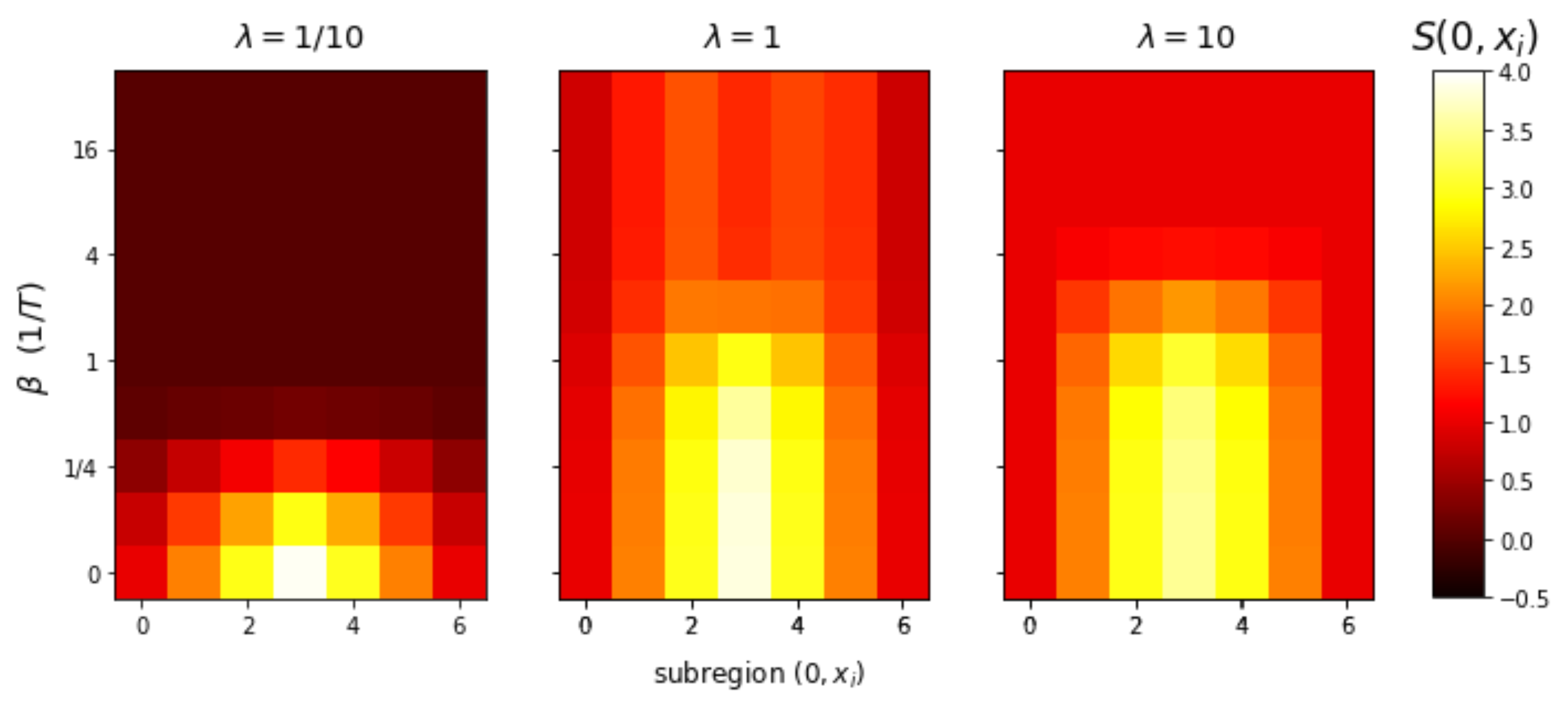}
  \end{center}
  \caption{Entanglement entropy of the thermofield-double state as a function of the inverse temperature $\beta = T^{-1}$, for various values of the transverse coupling $\lambda$. }
  \label{fig:TFD_Tall}
\end{figure}
\end{center}

In order to examine the scaling properties we compute the entanglement entropy of the $\TFD$ state. The entanglement entropy $S(A)$ of a subregion of space $A$, quantifies the amount by which the degrees of freedom in $A$ are correlated with degrees of freedom not contained in $A$. It is defined in terms of the reduced density matrix $\rho_A$, which is the partial trace of the full density matrix $\rho$ over the complement $A^c$ as in Eq \ref{eq:EE}. The degree of entanglement is measured by the degree to which the pure state density matrix $\rho$ becomes a mixed state $\rho_A$ under the partial trace.

\be
S(A) = - \Tr \left( \rho_A \log \rho_A \right) \qquad \qquad \rho_A = \Tr_{A^c} \, \rho
\label{eq:EE}
\ee
For the quantum XY chain with periodic boundary conditions, the entanglement entropy $S(\ell)$ corresponding to a subregion of length $\ell$ scales logarithmically at the quantum critical point corresponding to $\lambda = 1$. The analyical formula for $S(\ell)$ is given in Eq \ref{eq:Scrit} in terms of the central charge $c = 1/2$. Away from the critical point $S (\ell)$ scales roughly as a constant depending on the correlation length $\xi$, as shown in Eq \ref{eq:Scrit}
\be
S_{\lambda=1} (\ell) = \frac{c}{3} \log_2 \left( \frac{L}{\pi} \sin \frac{\ell \pi}{L} \right) + c_1 \qquad \qquad S_{\lambda \neq 1} (\ell) \sim \frac{c}{3} \log_2 \xi
\label{eq:Scrit}
\ee
%\be
%S_{\lambda \neq 1} (\ell) \sim \frac{c}{3} \log_2 \xi
%\label{eq:Snoncrit}
%\ee
In Fig \ref{fig:TFD_T1} we plot the entanglement entropy for the ground state and the thermo-field double at inverse temperature $\beta=1$, for various values of the transverse coupling $\lambda$. At the critical value $\lambda = 1$, we can see the characteristic logarithmic scaling of $S(\ell)$ in the ground state. Correspondingly we see that the logarithmic behavior is traded for a linear scaling in the thermofield-double, thus implying a local rescaling by an exponential factor. Fig \ref{fig:TFD_Tall} shows a plot of the entanglement entropy for the thermofield-double state as a function of the inverse temperature $\beta = T^{-1}$.

\section{Summary}
\label{sec:summary}

In this note we have explored algorithms for simulating simple integrable lattice models using an exact representation of the time evolution operator, without the need for a Trotterized approximation. We show that the complete control over the degrees of freedom afforded by these models allows one to perform a coarse-graining procedure that integrates over microscopic degrees of freedom, by taking a partial trace over momentum modes that lie above some ultraviolet cutoff $\Lambda$. We then described an algorithm for transforming the ground state of these theories to the highly entangled thermo-field double state, and verify that this is equivalent to a logarithmic re-scaling of position space coordinates. These methods could be made applicable to any system whose Hamiltonian is diagonalizable via two-local operations, by replacing the Bogoliubov transformations with a more complicated momentum-dependent scattering matrix. Although the relationship between the analogous procedures on the MERA network are unclear to this author presently, these circuits represent unitary operations for performing the analogous transformations. The simplest versions of these algorithms can be readily simulated on the noisy quantum computers available for public use today, which are capable of reliably entangling $\mathcal{O} ({\rm few})$ qubits. 

\newpage

\appendix

\section{Parameterization of Clifford Rotations}
\label{app:circuits}

Here we review a general construction for unitary gates that correspond to Clifford operations. The Clifford algebra can be represented on the space of $n$ qubits by a set of $2n$ matrices which are $2^n \times 2^n$ dimensional, known as the $\gamma^{(n)}$-matrices. On the space of $2$ qubits, the four $\gamma^{(2)}$-matrices have dimension $4 \times 4$ and are analogous to the Dirac matrices familiar to particle physicists. The defining property of the $\gamma$-matrices is the anti-commutation relation in Eq \ref{eq:gamma}. Additionally the commutator of the $\gamma$-matrices in Eq \ref{eq:lorentz} furnishes the spinorial representation of the generators for the Lie group $SO(2^n)$.

\be
\left\{ \gamma_i^{(n)}, \gamma_j^{(n)} \right\} = \delta_{i j}
\label{eq:gamma}
\ee

\be
\Sigma_{i j}^{(n)} = -\frac{i}{4} \Big[ \gamma_i^{(n)} , \gamma_j^{(n)} \Big]
\label{eq:lorentz}
\ee
An explicit form of the $\gamma$-matrices can be constructed iteratively. In this note we use a convention in which the 2-qubit $\gamma$-matrices take the form in Eq \ref{eq:gamma2} (left). Higher qubit representations can then be constructed through the given iterative procedure in Eq \ref{eq:gamma2} (right).
\be
\gamma^{(2)}_i = \begin{pmatrix}
I \otimes X \\
I \otimes Y \\
X \otimes Z \\
Y \otimes Z
\end{pmatrix} \qquad \qquad \gamma^{(n)}_i = \begin{pmatrix}
I \otimes I \otimes \gamma^{(n - 2)} \\[0.2cm]
\gamma^{(n - 2)} \otimes Z \otimes Z
\end{pmatrix}
\label{eq:gamma2}
\ee
From these $\gamma$-matrices we can construct a representation of the spinor states by defining a set of raising and lowering operators $c^{(n) \, \dagger}_j$ and $c^{(n)}_j$ as in Eq \ref{eq:gammaferm}. These operators obey the usual fermionic anticommutation relations shown in Eq \ref{eq:gammacomm}, where we have suppressed the dimensional index

\be
c_j = \frac{1}{2} \left( \gamma_{2 j - 1} + i \gamma_{2 j} \right)
\label{eq:gammaferm}
\ee

\be
\left\{ c_i^\dagger, c_j \right\} = \delta_{ij}  \qquad \qquad \left\{ c_i^\dagger , c_j^\dagger \right\} = \left\{ c_i, c_j \right\} = 0
\label{eq:gammacomm}
\ee
States in the spinorial representation of $SO(2^n)$ can be conveniently represented as a binary string $\ket{s_1 \, s_2 \, ... \, s_{n}}$ where $s_i \in (+, -)$. The ground state spinor $\ket{-}^{\otimes n}$ is defined as the state which is annihilated by all of the lowering operators $c_i \ket{-}^{\otimes n} = 0$ for all $i$. The $2^n$ dimensional spinor representation can then be constructed by acting on the ground state $\ket{-}^{\otimes n}$ in all possible ways with the $c_i^\dagger$ at most once each. For example on $n=5$ qubits
$$
c_4^\dagger c_3^\dagger c_1 ^\dagger \ket{-----} = \ket{+-++-}
$$

The orbit of $SO(2^n)$ on the $2^n$ dimensional Hilbert space of $n$ qubits can be traversed by successive products of $SO(4)$ rotations entangling two qubits. These $SO(4)$ rotations can be further decomposed via the isomorphism $SO(4) = SU(2)^+ \times SU(2)^-$. In terms of the spinorial generators of $SO(4)$ given in Eq \ref{eq:lorentz}, the $SU(2)^+ \times SU(2)^-$ generators are defined as in Eq \ref{eq:su2xsu2} and their representations in terms of the Pauli matrices is given in Eq \ref{eq:su2pm}.

\be
\Sigma^\pm_{\alpha} = \Sigma_{\alpha 4}^{(2)} \pm \frac{1}{2} \epsilon_{\alpha \beta \gamma 4} \Sigma_{\beta \gamma}^{(2)} \qquad \qquad \alpha = 1, 2, 3
\label{eq:su2xsu2}
\ee

\be
\Sigma^\pm_{X, Y, Z} = \frac{1}{2} \begin{pmatrix}
-X \otimes X \pm Y \otimes Y \\[0.2cm]
~Y \otimes X \pm X \otimes Y \\[0.2cm]
~Z \otimes I \pm I \otimes Z
\end{pmatrix}
\label{eq:su2pm}
\ee
All of the circuits used in this paper can be parameterized by the unitary $SU(2)^\pm$ rotations in \ref{eq:unitarypm} about their $X$, $Y$, and $Z$ axes. The generators of $SU(2)^+$ correspond to operations that flip two spins and are equivalent to the fermionic operators $c_i^\dagger c_j^\dagger$ and $c_i c_j$ which change the total fermion number by 2. The generators of $SU(2)^-$ correspond to operations that preserve the fermion number and are equivalent to $c_i^\dagger c_j$ and $c_i c_j^\dagger$.
\be
U_{X,Y,Z}^\pm (\theta) = e^{-\frac{i}{2}  \theta \Sigma_{X,Y,Z}^{\pm} }
\label{eq:unitarypm}
\ee
Here we show the textbook implementation \cite{Nielsen:2011:QCQ:1972505} of these $SU(2)^\pm$ rotations in terms of the standard $H$, $S$ and CNOT gates. Although more efficient constructions using fewer CNOTs are also known, the textbook implementation is shown for its pedagogical transparency.

\begin{center}
\begin{figure}[ht]
\begin{tikzcd}[row sep={1cm,between origins}]
& \gate[wires=2]{~U_{X}^\pm (\theta)~} & \qw \\ 
& \hphantom{wide label} & \qw
\end{tikzcd}
=
\begin{tikzcd}[row sep={1cm,between origins}]
& \gate{{\SG}} & \gate{\HG} & \ctrl{1} & \qw & \ctrl{1}  & \gate{\HG} & \gate{\SG^\dagger} & \gate{\HG} & \ctrl{1} & \qw & \ctrl{1} & \gate{\HG} & \qw \\ 
& \gate{{\SG}} & \gate{\HG} & \targ{ } & \gate{e^{\mp \frac{i}{4}  \theta Z}} & \targ{ } & \gate{\HG} & \gate{\SG^\dagger} & \gate{\HG} & \targ{ } & \gate{e^{+ \frac{i}{4}  \theta Z}} & \targ{ } & \gate{\HG} & \qw
\end{tikzcd}\\
\vspace{1cm}

\begin{tikzcd}[row sep={1cm,between origins}]
& \gate[wires=2]{~U_{Y}^\pm (\theta)~} & \qw \\ 
& \hphantom{wide label} & \qw
\end{tikzcd}
=
\begin{tikzcd}[row sep={1cm,between origins}]
& \qw & \gate{\HG} & \ctrl{1} & \qw & \ctrl{1}  & \gate{\HG} & \gate{\SG} & \gate{\HG} & \ctrl{1} & \qw & \ctrl{1} & \gate{\HG} & \gate{{\SG^\dagger}} & \qw \\ 
& \gate{{\SG}} & \gate{\HG} & \targ{ } & \gate{e^{\mp \frac{i}{4}  \theta Z}} & \targ{ } & \gate{\HG} & \gate{\SG^\dagger} & \gate{\HG} & \targ{ } & \gate{e^{- \frac{i}{4}  \theta Z}} & \targ{ } & \gate{\HG} & \qw & \qw
\end{tikzcd}
\caption{The textbook implementation of $SU(2)^\pm$ rotations in terms of CNOT gates. These circuits all work by implementing an entangling $e^{-\frac{i}{2} \theta Z \otimes Z}$ rotation, followed by an appropriate conjugation by Hadamard and $S$ gates.}
\label{diag:ULR}
\end{figure}
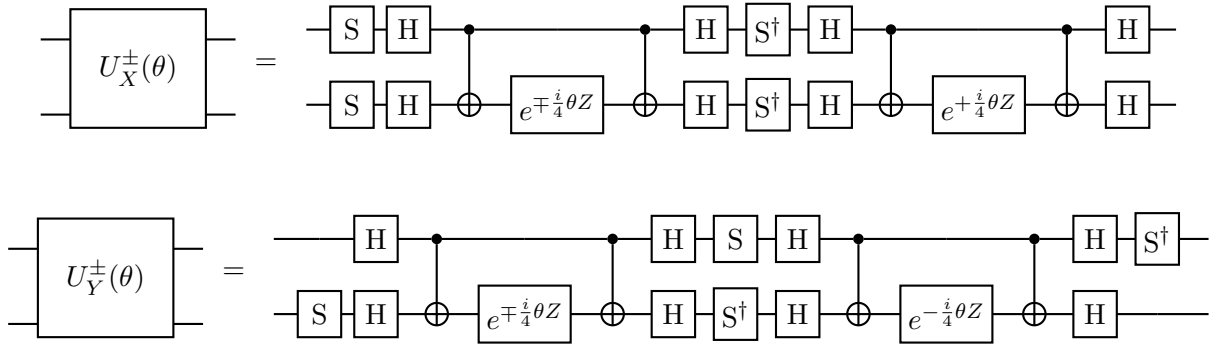
\end{center}

\begin{center}
\begin{figure}[ht]
\begin{tikzcd}[row sep={0.75cm,between origins}]
& \gate[wires=2]{~U_{Z}^\pm (\theta)~} & \qw \\ 
& \hphantom{wide label} & \qw
\end{tikzcd}
=
\begin{tikzcd}[row sep={1cm,between origins}]
& \gate{e^{- \frac{i}{4} \theta Z}} & \qw  \\ 
& \gate{e^{\mp \frac{i}{4} \theta Z}} & \qw
\end{tikzcd}
\caption{An $SU(2)^\pm$ rotation about the $Z$ direction can be performed simply by applying single qubit phase rotations.}
\label{diag:ULR}
\end{figure}
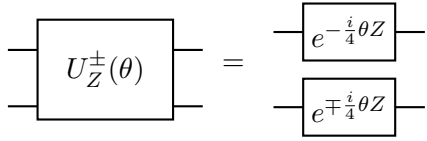
\end{center}

\section{Explicit Circuit Constructions}
\label{app:circuits}

In terms of the $SU(2)^+ \times SU(2)^-$ rotations described in the previous section, the two-qubit Bogoliubov gates are simply given by an $SU(2)^+$ rotation about the $X$ direction by the angle $\theta_p$ as defined in Eq \ref{eq:Bog2}.

\begin{center}
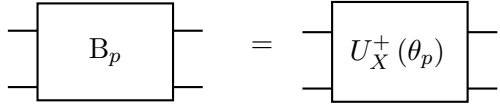
\begin{figure}[ht]
\begin{tikzcd}[row sep={0.75cm,between origins}]
& \gate[wires=2]{~{\Bog}_p ~} & \qw \\ 
& \hphantom{wide label} & \qw
\end{tikzcd}
\quad =
\begin{tikzcd}[row sep={0.75cm,between origins}]
& \gate[wires=2]{~U_{X}^+ \left( \theta_p \right)~} & \qw & \\ 
& \hphantom{wide label} & \qw &
\end{tikzcd}
\caption{A circuit representation of the two-qubit Bogoliubov transformation. It is equivalent to an $SU(2)^+$ rotation about the X direction by the Bogoliubov angle $\theta_p$}
\label{diag:CliffBog}
\end{figure}
\end{center}
\vspace{-0.5cm}
Next we describe the procedure for implementing the Fourier transform in terms of two-qubit local gates. Although more efficient implementations of the fermionic Fourier transform are known \cite{Jiang:2017pyp}, here we describe the original construction given in \cite{Verstraete:2008qpa} due to its pedagogical amenability. The Fourier transform on fermions can be performed by iteratively applying a sequence of two-body local gates ${\rm F}_p$ combined with two-body ${\rm SWAP}$ operations. On four qubits this iterative process may be represented as

\begin{align}
\tilde{c}_p &= \frac{1}{4} \displaystyle\sum_{j = 1}^4 e^{2 \pi i p j} c_j \\
&= c_1 + e^{2 \pi i (2 p)} c_3 + e^{2 \pi i (2 p)} \left( c_2 + e^{2 \pi i (2 p)} c_4 \right)
\end{align}
The two-body Fourier gate which implements one step of this operation can be written in terms of a phase rotation on one qubit ${\rm R} (\theta)$ followed by an $SU(2)^-$ rotation about the $Y$ direction by $\pi/4$ as shown in Fig \ref{diag:CliffFT}.
\begin{center}
\begin{figure}[ht]
\begin{tikzcd}[row sep={0.75cm,between origins}]
& \gate[wires=2]{~{\rm R}(\theta)~} & \qw \\ 
& & \qw
\end{tikzcd}
\quad =
\begin{tikzcd}[row sep={0.75cm,between origins}]
& \gate[wires=2]{~e^{+\frac{i}{2} \theta}~} & \gate[wires=2]{~U^Z_+ (\theta)~} & \gate[wires=2]{~U^Z_- (\theta)~} & \qw \\ 
& & & \qw & \qw
\end{tikzcd}\\
\vspace{1cm}

\begin{tikzcd}[row sep={0.75cm,between origins}]
& \gate[wires=2]{~{\rm F}_p ~} & \qw \\ 
& \hphantom{wide label} & \qw
\end{tikzcd}
\quad =
\begin{tikzcd}[row sep={0.75cm,between origins}]
& \gate[wires=2]{~U_{Y}^- \left( \frac{\pi}{2} \right)~} & \gate[wires=2]{{\rm R} ( 2 \pi p+\pi)} & \qw & \\ 
&  &  & \qw &
\end{tikzcd}
\caption{A Fourier transform on $n$ qubits can be performed by iterative applications of the two-body Fourier gate ${\rm F}_p$, which implements a relative phase rotation between two qubits.}
\label{diag:CliffFT}
\end{figure}
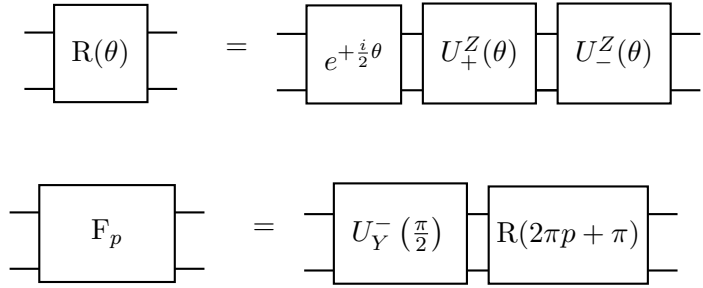
\end{center}

The constraint of two-body locality in this note relies heavily on SWAP operations that exchange the physical location of various fermionic modes. The two qubit SWAP gate can be expressed in terms of an $SU(2)^-$ rotation about the $Y$ direction, followed by a phase rotation by $\pi$ to account for the fermionic statistics. Using the two-body SWAP gate, momentum modes can be arbitrarily arranged in the physical space of qubits as shown in Fig \ref{diag:fSWAP}

\begin{center}
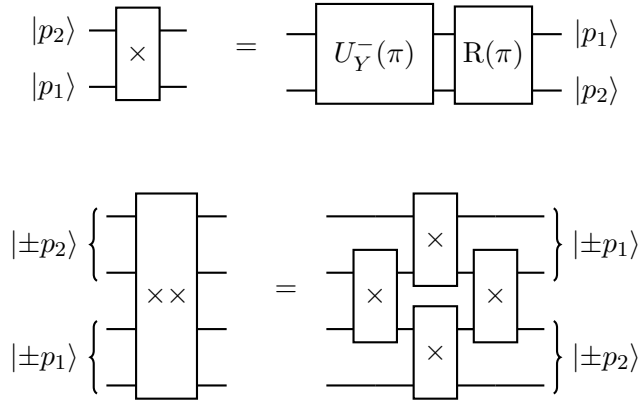
\begin{figure}[ht]
\begin{tikzcd}[row sep={0.75cm,between origins}]
\lstick{\ket{p_2}} & \gate[wires=2]{\times } & \qw \\ 
\lstick{\ket{p_1}} & & \qw 
\end{tikzcd}
\quad =
\begin{tikzcd}[row sep={0.75cm,between origins}]
& \gate[wires=2]{~U_{Y}^- (\pi)~} & \gate[wires=2]{{\rm R}(\pi) } & \qw \rstick{\ket{p_1}} \\ 
& & & \qw \rstick{\ket{p_2}}
\end{tikzcd}\\
\vspace{1cm}

\begin{tikzcd}[row sep={0.75cm,between origins}]
\lstick[wires=2]{\ket{\pm p_2}} & \gate[wires=4]{\times \times } & \qw \\ 
& & \qw \\
\lstick[wires=2]{\ket{\pm p_1}} & & \qw \\
& & \qw
\end{tikzcd}
\quad =
\begin{tikzcd}[row sep={0.75cm,between origins}]
& \qw & \gate[wires=2]{\times } & \qw & \qw \rstick[wires=2]{\ket{\pm p_1}} \\ 
& \gate[wires=2]{\times } & & \gate[wires=2]{\times } & \qw \\
& & \gate[wires=2]{\times } & \qw & \qw \rstick[wires=2]{\ket{\pm p_2}} \\
& \qw & & \qw & \qw
\end{tikzcd}
\caption{A fermionic SWAP operation that exchanges the physical location of encoded states between two qubits. Higher qubit SWAP gates can be constructed from the two-qubit SWAP.}
\label{diag:fSWAP}
\end{figure}
\end{center}
In Fig \ref{diag:FTcircuit} we show a circuit that implements the Fourier transform on $n=8$ qubits used in these simulations. The output of this transform localizes pairs of momentum with opposite sign grouping together even (odd) modes on the upper (lower) half of the qubits. In order to implement the partial trace over high momentum modes in a local fashion, the sequence of SWAP gates shown in Fig \ref{diag:SWAPcircuit} can be used to localize the high momentum modes on the lower half of the qubits.
\begin{center}
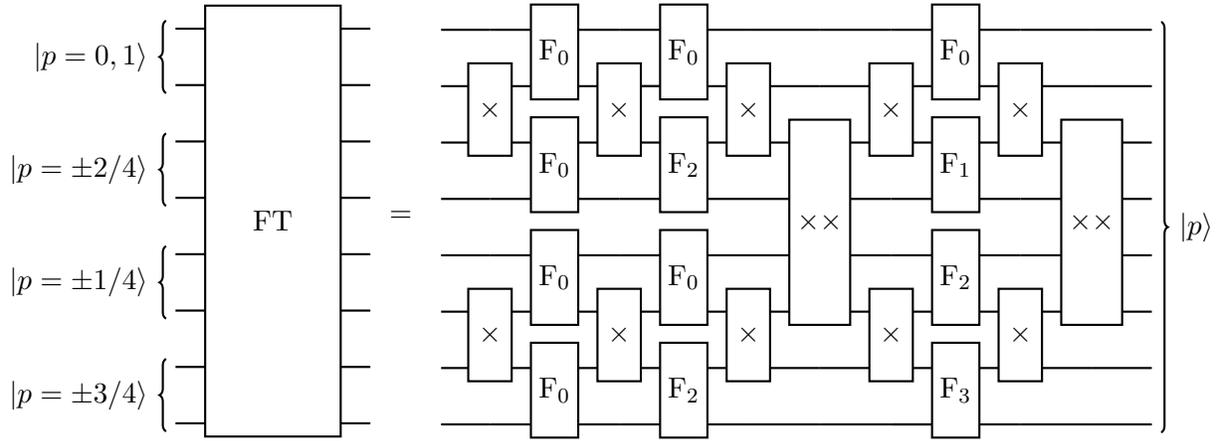
\begin{figure}[ht]
\begin{tikzcd}[row sep={0.75cm,between origins}]
\lstick[wires=2]{$\ket{p=0,1}$} & \gate[wires=8]{\rm FT} & \qw \\ 
 & & \qw \\ 
\lstick[wires=2]{$\ket{p=\pm 2/4}$} & & \qw \\
& & \qw \\
\lstick[wires=2]{$\ket{p=\pm 1/4}$} & & \qw \\
& & \qw \\
\lstick[wires=2]{$\ket{p=\pm 3/4}$} & & \qw \\
 & \hphantom{wide label} & \qw
\end{tikzcd}
=
\begin{tikzcd}[row sep={0.75cm,between origins}]
& \qw & \gate[wires=2]{{\rm F}_0 } & \qw & \gate[wires=2]{{\rm F}_0 } & \qw & \qw & \qw & \gate[wires=2]{{\rm F}_0 } & \qw & \qw & \qw \rstick[wires=8]{\ket{p}} \\ 
& \gate[wires=2]{\times } & & \gate[wires=2]{\times } & \qw & \gate[wires=2]{\times } & \qw & \gate[wires=2]{\times } & & \gate[wires=2]{\times } & \qw & \qw \\
& & \gate[wires=2]{{\rm F}_0 } & & \gate[wires=2]{{\rm F}_2 } & & \gate[wires=4]{\times \times } & & \gate[wires=2]{{\rm F}_1 } & & \gate[wires=4]{\times \times } & \qw \\
& \qw & & \qw & & \qw & & \qw & & \qw &  & \qw \\
& \qw & \gate[wires=2]{{\rm F}_0 } & \qw & \gate[wires=2]{{\rm F}_0 } & \qw & & \qw & \gate[wires=2]{{\rm F}_2 } & \qw &  & \qw \\
& \gate[wires=2]{\times } & & \gate[wires=2]{\times } & & \gate[wires=2]{\times } & & \gate[wires=2]{\times } & & \gate[wires=2]{\times } &  & \qw \\
& & \gate[wires=2]{{\rm F}_0 } & &\gate[wires=2]{{\rm F}_2 } & & \qw & & \gate[wires=2]{{\rm F}_3 } & & \qw & \qw \\
& \qw & & \qw & & \qw & \qw & \qw & \qw & \qw & \qw & \qw
\end{tikzcd}
\caption{A circuit representation of the fermionic Fourier transform on a system of $n=8$ qubits. The circuit proceeds by performing relative phase rotations on pairs of qubits, facilitated by SWAP operations that exchange the pairs to be rotated.}
\label{diag:FTcircuit}
\end{figure}
\end{center}

\begin{center}
\begin{figure}[ht]
\begin{tikzcd}[row sep={0.75cm,between origins}]
\lstick[wires=4]{$\ket{p<2/4}$} & \gate[wires=8]{\rm SWAP} & \qw \\ 
 & & \qw \\ 
& & \qw \\
& & \qw \\
\lstick[wires=4]{$\ket{p>2/4}$} & & \qw \\
& & \qw \\
& & \qw \\
 & \hphantom{wide label} & \qw
\end{tikzcd}
=
\begin{tikzcd}[row sep={0.75cm,between origins}]
& \qw & \qw & \qw \rstick[wires=2]{$\ket{p=0,1}$} \\ 
& \qw & \gate[wires=2]{\times } & \qw \\
& \gate[wires=4]{\times \times } & & \qw \rstick[wires=2]{$\ket{p=\pm 2/4}$} \\
& & \qw & \qw \\
& & \qw & \qw \rstick[wires=2]{$\ket{p=\pm 1/4}$} \\
& & \qw & \qw \\
& \qw & \qw & \qw \rstick[wires=2]{$\ket{p=\pm 3/4}$} \\
& \qw & \qw & \qw
\end{tikzcd}
\caption{A sequence of SWAP operations that localize high and low momentum modes onto neighboring qubits. Given any arrangement of momentum space modes output by a Fourier transform, a sequence of SWAP gates exists to localize the modes of interest.}
\label{diag:SWAPcircuit}
\end{figure}
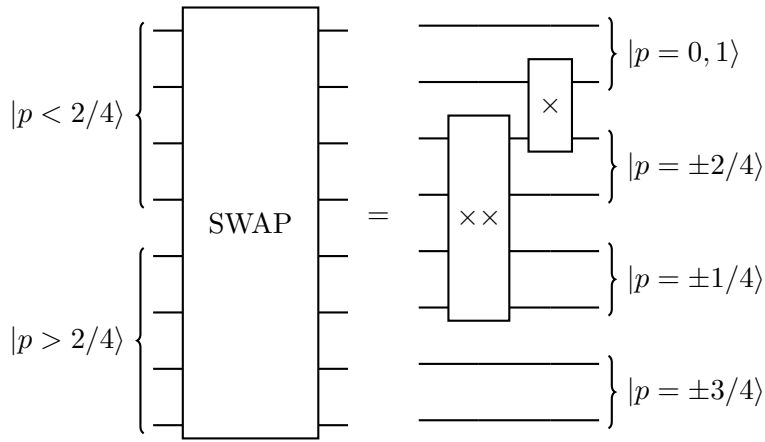
\end{center}

\newpage


\begin{thebibliography}{48}%
\makeatletter
\providecommand \@ifxundefined [1]{%
 \@ifx{#1\undefined}
}%
\providecommand \@ifnum [1]{%
 \ifnum #1\expandafter \@firstoftwo
 \else \expandafter \@secondoftwo
 \fi
}%
\providecommand \@ifx [1]{%
 \ifx #1\expandafter \@firstoftwo
 \else \expandafter \@secondoftwo
 \fi
}%
\providecommand \natexlab [1]{#1}%
\providecommand \enquote  [1]{``#1''}%
\providecommand \bibnamefont  [1]{#1}%
\providecommand \bibfnamefont [1]{#1}%
\providecommand \citenamefont [1]{#1}%
\providecommand \href@noop [0]{\@secondoftwo}%
\providecommand \href [0]{\begingroup \@sanitize@url \@href}%
\providecommand \@href[1]{\@@startlink{#1}\@@href}%
\providecommand \@@href[1]{\endgroup#1\@@endlink}%
\providecommand \@sanitize@url [0]{\catcode `\\12\catcode `\$12\catcode
  `\&12\catcode `\#12\catcode `\^12\catcode `\_12\catcode `\%12\relax}%
\providecommand \@@startlink[1]{}%
\providecommand \@@endlink[0]{}%
\providecommand \url  [0]{\begingroup\@sanitize@url \@url }%
\providecommand \@url [1]{\endgroup\@href {#1}{\urlprefix }}%
\providecommand \urlprefix  [0]{URL }%
\providecommand \Eprint [0]{\href }%
\providecommand \doibase [0]{http://dx.doi.org/}%
\providecommand \selectlanguage [0]{\@gobble}%
\providecommand \bibinfo  [0]{\@secondoftwo}%
\providecommand \bibfield  [0]{\@secondoftwo}%
\providecommand \translation [1]{[#1]}%
\providecommand \BibitemOpen [0]{}%
\providecommand \bibitemStop [0]{}%
\providecommand \bibitemNoStop [0]{.\EOS\space}%
\providecommand \EOS [0]{\spacefactor3000\relax}%
\providecommand \BibitemShut  [1]{\csname bibitem#1\endcsname}%
\let\auto@bib@innerbib\@empty
%</preamble>
\bibitem [{\citenamefont {Nielsen}\ and\ \citenamefont
  {Chuang}(2011)}]{Nielsen:2011:QCQ:1972505}%
  \BibitemOpen
  \bibfield  {author} {\bibinfo {author} {\bibfnamefont {M.~A.}\ \bibnamefont
  {Nielsen}}\ and\ \bibinfo {author} {\bibfnamefont {I.~L.}\ \bibnamefont
  {Chuang}},\ }\href@noop {} {\emph {\bibinfo {title} {Quantum Computation and
  Quantum Information: 10th Anniversary Edi tion}}},\ \bibinfo {edition}
  {10th}\ ed.\ (\bibinfo  {publisher} {Cambridge University Press},\ \bibinfo
  {address} {New York, NY, USA},\ \bibinfo {year} {2011})\BibitemShut {NoStop}%
\bibitem [{\citenamefont {Maldacena}(1999)}]{Maldacena:1997re}%
  \BibitemOpen
  \bibfield  {author} {\bibinfo {author} {\bibfnamefont {J.~M.}\ \bibnamefont
  {Maldacena}},\ }\href {\doibase 10.1023/A:1026654312961,
  10.4310/ATMP.1998.v2.n2.a1} {\bibfield  {journal} {\bibinfo  {journal} {Int.
  J. Theor. Phys.}\ }\textbf {\bibinfo {volume} {38}},\ \bibinfo {pages} {1113}
  (\bibinfo {year} {1999})},\ \bibinfo {note} {[Adv. Theor. Math.
  Phys.2,231(1998)]},\ \Eprint {http://arxiv.org/abs/hep-th/9711200}
  {arXiv:hep-th/9711200 [hep-th]} \BibitemShut {NoStop}%
%%CITATION = HEP-TH/9711200;%%
\bibitem [{\citenamefont {Takahashi}\ and\ \citenamefont
  {Umezawa}(1996)}]{Takahashi:1996zn}%
  \BibitemOpen
  \bibfield  {author} {\bibinfo {author} {\bibfnamefont {Y.}~\bibnamefont
  {Takahashi}}\ and\ \bibinfo {author} {\bibfnamefont {H.}~\bibnamefont
  {Umezawa}},\ }\href {\doibase 10.1142/S0217979296000817} {\bibfield
  {journal} {\bibinfo  {journal} {Int. J. Mod. Phys.}\ }\textbf {\bibinfo
  {volume} {B10}},\ \bibinfo {pages} {1755} (\bibinfo {year}
  {1996})}\BibitemShut {NoStop}%
%%CITATION = IMPAE,B10,1755;%%
\bibitem [{\citenamefont {{Suzuki}}(1990)}]{1990PhLA..146..319S}%
  \BibitemOpen
  \bibfield  {author} {\bibinfo {author} {\bibfnamefont {M.}~\bibnamefont
  {{Suzuki}}},\ }\href {\doibase 10.1016/0375-9601(90)90962-N} {\bibfield
  {journal} {\bibinfo  {journal} {Physics Letters A}\ }\textbf {\bibinfo
  {volume} {146}},\ \bibinfo {pages} {319} (\bibinfo {year}
  {1990})}\BibitemShut {NoStop}%
\bibitem [{\citenamefont {{Suzuki}}(1991)}]{1991JMP....32..400S}%
  \BibitemOpen
  \bibfield  {author} {\bibinfo {author} {\bibfnamefont {M.}~\bibnamefont
  {{Suzuki}}},\ }\href@noop {} {\bibfield  {journal} {\bibinfo  {journal}
  {Journal of Mathematical Physics}\ } (\bibinfo {year} {1991})}\BibitemShut
  {NoStop}%
\bibitem [{\citenamefont {Hawking}()}]{Hawking:1974sw}%
  \BibitemOpen
  \bibfield  {author} {\bibinfo {author} {\bibfnamefont {S.~W.}\ \bibnamefont
  {Hawking}},\ }\href@noop {} {\ }\BibitemShut {NoStop}%
\bibitem [{\citenamefont {Page}(1993)}]{Page:1993df}%
  \BibitemOpen
  \bibfield  {author} {\bibinfo {author} {\bibfnamefont {D.~N.}\ \bibnamefont
  {Page}},\ }\href {\doibase 10.1103/PhysRevLett.71.1291} {\bibfield  {journal}
  {\bibinfo  {journal} {Phys. Rev. Lett.}\ }\textbf {\bibinfo {volume} {71}},\
  \bibinfo {pages} {1291} (\bibinfo {year} {1993})},\ \Eprint
  {http://arxiv.org/abs/gr-qc/9305007} {arXiv:gr-qc/9305007 [gr-qc]}
  \BibitemShut {NoStop}%
%%CITATION = GR-QC/9305007;%%
\bibitem [{\citenamefont {Susskind}\ \emph {et~al.}(1993)\citenamefont
  {Susskind}, \citenamefont {Thorlacius},\ and\ \citenamefont
  {Uglum}}]{Susskind:1993if}%
  \BibitemOpen
  \bibfield  {author} {\bibinfo {author} {\bibfnamefont {L.}~\bibnamefont
  {Susskind}}, \bibinfo {author} {\bibfnamefont {L.}~\bibnamefont
  {Thorlacius}}, \ and\ \bibinfo {author} {\bibfnamefont {J.}~\bibnamefont
  {Uglum}},\ }\href {\doibase 10.1103/PhysRevD.48.3743} {\bibfield  {journal}
  {\bibinfo  {journal} {Phys. Rev.}\ }\textbf {\bibinfo {volume} {D48}},\
  \bibinfo {pages} {3743} (\bibinfo {year} {1993})},\ \Eprint
  {http://arxiv.org/abs/hep-th/9306069} {arXiv:hep-th/9306069 [hep-th]}
  \BibitemShut {NoStop}%
%%CITATION = HEP-TH/9306069;%%
\bibitem [{\citenamefont {Hayden}\ and\ \citenamefont
  {Preskill}(2007)}]{Hayden:2007cs}%
  \BibitemOpen
  \bibfield  {author} {\bibinfo {author} {\bibfnamefont {P.}~\bibnamefont
  {Hayden}}\ and\ \bibinfo {author} {\bibfnamefont {J.}~\bibnamefont
  {Preskill}},\ }\href {\doibase 10.1088/1126-6708/2007/09/120} {\bibfield
  {journal} {\bibinfo  {journal} {JHEP}\ }\textbf {\bibinfo {volume} {09}},\
  \bibinfo {pages} {120} (\bibinfo {year} {2007})},\ \Eprint
  {http://arxiv.org/abs/0708.4025} {arXiv:0708.4025 [hep-th]} \BibitemShut
  {NoStop}%
%%CITATION = ARXIV:0708.4025;%%
\bibitem [{\citenamefont {Sekino}\ and\ \citenamefont
  {Susskind}(2008)}]{Sekino:2008he}%
  \BibitemOpen
  \bibfield  {author} {\bibinfo {author} {\bibfnamefont {Y.}~\bibnamefont
  {Sekino}}\ and\ \bibinfo {author} {\bibfnamefont {L.}~\bibnamefont
  {Susskind}},\ }\href {\doibase 10.1088/1126-6708/2008/10/065} {\bibfield
  {journal} {\bibinfo  {journal} {JHEP}\ }\textbf {\bibinfo {volume} {10}},\
  \bibinfo {pages} {065} (\bibinfo {year} {2008})},\ \Eprint
  {http://arxiv.org/abs/0808.2096} {arXiv:0808.2096 [hep-th]} \BibitemShut
  {NoStop}%
%%CITATION = ARXIV:0808.2096;%%
\bibitem [{\citenamefont {Susskind}(2016)}]{Susskind:2014rva}%
  \BibitemOpen
  \bibfield  {author} {\bibinfo {author} {\bibfnamefont {L.}~\bibnamefont
  {Susskind}},\ }\href {\doibase 10.1002/prop.201500093,
  10.1002/prop.201500092} {\bibfield  {journal} {\bibinfo  {journal} {Fortsch.
  Phys.}\ }\textbf {\bibinfo {volume} {64}},\ \bibinfo {pages} {44} (\bibinfo
  {year} {2016})},\ \bibinfo {note} {[Fortsch. Phys.64,24(2016)]},\ \Eprint
  {http://arxiv.org/abs/1403.5695} {arXiv:1403.5695 [hep-th]} \BibitemShut
  {NoStop}%
%%CITATION = ARXIV:1403.5695;%%
\bibitem [{\citenamefont {Susskind}(2018)}]{Susskind:2018fmx}%
  \BibitemOpen
  \bibfield  {author} {\bibinfo {author} {\bibfnamefont {L.}~\bibnamefont
  {Susskind}},\ }\href@noop {} {\  (\bibinfo {year} {2018})},\ \Eprint
  {http://arxiv.org/abs/1802.02175} {arXiv:1802.02175 [hep-th]} \BibitemShut
  {NoStop}%
%%CITATION = ARXIV:1802.02175;%%
\bibitem [{\citenamefont {Almheiri}\ \emph {et~al.}(2013)\citenamefont
  {Almheiri}, \citenamefont {Marolf}, \citenamefont {Polchinski},\ and\
  \citenamefont {Sully}}]{Almheiri:2012rt}%
  \BibitemOpen
  \bibfield  {author} {\bibinfo {author} {\bibfnamefont {A.}~\bibnamefont
  {Almheiri}}, \bibinfo {author} {\bibfnamefont {D.}~\bibnamefont {Marolf}},
  \bibinfo {author} {\bibfnamefont {J.}~\bibnamefont {Polchinski}}, \ and\
  \bibinfo {author} {\bibfnamefont {J.}~\bibnamefont {Sully}},\ }\href
  {\doibase 10.1007/JHEP02(2013)062} {\bibfield  {journal} {\bibinfo  {journal}
  {JHEP}\ }\textbf {\bibinfo {volume} {02}},\ \bibinfo {pages} {062} (\bibinfo
  {year} {2013})},\ \Eprint {http://arxiv.org/abs/1207.3123} {arXiv:1207.3123
  [hep-th]} \BibitemShut {NoStop}%
%%CITATION = ARXIV:1207.3123;%%
\bibitem [{\citenamefont {Susskind}(2012)}]{Susskind:2012rm}%
  \BibitemOpen
  \bibfield  {author} {\bibinfo {author} {\bibfnamefont {L.}~\bibnamefont
  {Susskind}},\ }\href@noop {} {\  (\bibinfo {year} {2012})},\ \Eprint
  {http://arxiv.org/abs/1208.3445} {arXiv:1208.3445 [hep-th]} \BibitemShut
  {NoStop}%
%%CITATION = ARXIV:1208.3445;%%
\bibitem [{\citenamefont {Maldacena}\ and\ \citenamefont
  {Susskind}(2013)}]{Maldacena:2013xja}%
  \BibitemOpen
  \bibfield  {author} {\bibinfo {author} {\bibfnamefont {J.}~\bibnamefont
  {Maldacena}}\ and\ \bibinfo {author} {\bibfnamefont {L.}~\bibnamefont
  {Susskind}},\ }\href {\doibase 10.1002/prop.201300020} {\bibfield  {journal}
  {\bibinfo  {journal} {Fortsch. Phys.}\ }\textbf {\bibinfo {volume} {61}},\
  \bibinfo {pages} {781} (\bibinfo {year} {2013})},\ \Eprint
  {http://arxiv.org/abs/1306.0533} {arXiv:1306.0533 [hep-th]} \BibitemShut
  {NoStop}%
%%CITATION = ARXIV:1306.0533;%%
\bibitem [{\citenamefont {Harlow}\ and\ \citenamefont
  {Hayden}(2013)}]{Harlow:2013tf}%
  \BibitemOpen
  \bibfield  {author} {\bibinfo {author} {\bibfnamefont {D.}~\bibnamefont
  {Harlow}}\ and\ \bibinfo {author} {\bibfnamefont {P.}~\bibnamefont
  {Hayden}},\ }\href {\doibase 10.1007/JHEP06(2013)085} {\bibfield  {journal}
  {\bibinfo  {journal} {JHEP}\ }\textbf {\bibinfo {volume} {06}},\ \bibinfo
  {pages} {085} (\bibinfo {year} {2013})},\ \Eprint
  {http://arxiv.org/abs/1301.4504} {arXiv:1301.4504 [hep-th]} \BibitemShut
  {NoStop}%
%%CITATION = ARXIV:1301.4504;%%
\bibitem [{\citenamefont {Ryu}\ and\ \citenamefont
  {Takayanagi}(2006{\natexlab{a}})}]{Ryu:2006bv}%
  \BibitemOpen
  \bibfield  {author} {\bibinfo {author} {\bibfnamefont {S.}~\bibnamefont
  {Ryu}}\ and\ \bibinfo {author} {\bibfnamefont {T.}~\bibnamefont
  {Takayanagi}},\ }\href {\doibase 10.1103/PhysRevLett.96.181602} {\bibfield
  {journal} {\bibinfo  {journal} {Phys. Rev. Lett.}\ }\textbf {\bibinfo
  {volume} {96}},\ \bibinfo {pages} {181602} (\bibinfo {year}
  {2006}{\natexlab{a}})},\ \Eprint {http://arxiv.org/abs/hep-th/0603001}
  {arXiv:hep-th/0603001 [hep-th]} \BibitemShut {NoStop}%
%%CITATION = HEP-TH/0603001;%%
\bibitem [{\citenamefont {Ryu}\ and\ \citenamefont
  {Takayanagi}(2006{\natexlab{b}})}]{Ryu:2006ef}%
  \BibitemOpen
  \bibfield  {author} {\bibinfo {author} {\bibfnamefont {S.}~\bibnamefont
  {Ryu}}\ and\ \bibinfo {author} {\bibfnamefont {T.}~\bibnamefont
  {Takayanagi}},\ }\href {\doibase 10.1088/1126-6708/2006/08/045} {\bibfield
  {journal} {\bibinfo  {journal} {JHEP}\ }\textbf {\bibinfo {volume} {08}},\
  \bibinfo {pages} {045} (\bibinfo {year} {2006}{\natexlab{b}})},\ \Eprint
  {http://arxiv.org/abs/hep-th/0605073} {arXiv:hep-th/0605073 [hep-th]}
  \BibitemShut {NoStop}%
%%CITATION = HEP-TH/0605073;%%
\bibitem [{\citenamefont {Pastawski}\ \emph {et~al.}(2015)\citenamefont
  {Pastawski}, \citenamefont {Yoshida}, \citenamefont {Harlow},\ and\
  \citenamefont {Preskill}}]{Pastawski:2015qua}%
  \BibitemOpen
  \bibfield  {author} {\bibinfo {author} {\bibfnamefont {F.}~\bibnamefont
  {Pastawski}}, \bibinfo {author} {\bibfnamefont {B.}~\bibnamefont {Yoshida}},
  \bibinfo {author} {\bibfnamefont {D.}~\bibnamefont {Harlow}}, \ and\ \bibinfo
  {author} {\bibfnamefont {J.}~\bibnamefont {Preskill}},\ }\href {\doibase
  10.1007/JHEP06(2015)149} {\bibfield  {journal} {\bibinfo  {journal} {JHEP}\
  }\textbf {\bibinfo {volume} {06}},\ \bibinfo {pages} {149} (\bibinfo {year}
  {2015})},\ \Eprint {http://arxiv.org/abs/1503.06237} {arXiv:1503.06237
  [hep-th]} \BibitemShut {NoStop}%
%%CITATION = ARXIV:1503.06237;%%
\bibitem [{\citenamefont {Harlow}(2018)}]{Harlow:2018fse}%
  \BibitemOpen
  \bibfield  {author} {\bibinfo {author} {\bibfnamefont {D.}~\bibnamefont
  {Harlow}},\ }\bibfield  {booktitle} {\emph {\bibinfo {booktitle}
  {{Proceedings, Theoretical Advanced Study Institute (TASI 2017)}}},\ }\href
  {\doibase 10.22323/1.305.0002} {\bibfield  {journal} {\bibinfo  {journal}
  {PoS}\ }\textbf {\bibinfo {volume} {TASI2017}},\ \bibinfo {pages} {002}
  (\bibinfo {year} {2018})},\ \Eprint {http://arxiv.org/abs/1802.01040}
  {arXiv:1802.01040 [hep-th]} \BibitemShut {NoStop}%
%%CITATION = ARXIV:1802.01040;%%
\bibitem [{\citenamefont {Vidal}(2004)}]{Vidal:2003lvx}%
  \BibitemOpen
  \bibfield  {author} {\bibinfo {author} {\bibfnamefont {G.}~\bibnamefont
  {Vidal}},\ }\href {\doibase 10.1103/PhysRevLett.93.040502} {\bibfield
  {journal} {\bibinfo  {journal} {Phys. Rev. Lett.}\ }\textbf {\bibinfo
  {volume} {93}},\ \bibinfo {pages} {040502} (\bibinfo {year} {2004})},\
  \Eprint {http://arxiv.org/abs/quant-ph/0310089} {arXiv:quant-ph/0310089
  [quant-ph]} \BibitemShut {NoStop}%
%%CITATION = QUANT-PH/0310089;%%
\bibitem [{\citenamefont {Vidal}(2007)}]{Vidal:2007hda}%
  \BibitemOpen
  \bibfield  {author} {\bibinfo {author} {\bibfnamefont {G.}~\bibnamefont
  {Vidal}},\ }\href {\doibase 10.1103/PhysRevLett.99.220405} {\bibfield
  {journal} {\bibinfo  {journal} {Phys. Rev. Lett.}\ }\textbf {\bibinfo
  {volume} {99}},\ \bibinfo {pages} {220405} (\bibinfo {year} {2007})},\
  \Eprint {http://arxiv.org/abs/cond-mat/0512165} {arXiv:cond-mat/0512165
  [cond-mat]} \BibitemShut {NoStop}%
%%CITATION = COND-MAT/0512165;%%
\bibitem [{\citenamefont {Evenbly}\ and\ \citenamefont
  {Vidal}(2009)}]{Evenbly:2007hxg}%
  \BibitemOpen
  \bibfield  {author} {\bibinfo {author} {\bibfnamefont {G.}~\bibnamefont
  {Evenbly}}\ and\ \bibinfo {author} {\bibfnamefont {G.}~\bibnamefont
  {Vidal}},\ }\href {\doibase 10.1103/PhysRevB.79.144108} {\bibfield  {journal}
  {\bibinfo  {journal} {Phys. Rev.}\ }\textbf {\bibinfo {volume} {B79}},\
  \bibinfo {pages} {144108} (\bibinfo {year} {2009})},\ \Eprint
  {http://arxiv.org/abs/0707.1454} {arXiv:0707.1454 [cond-mat.str-el]}
  \BibitemShut {NoStop}%
%%CITATION = ARXIV:0707.1454;%%
\bibitem [{\citenamefont {Swingle}(2012)}]{Swingle:2009bg}%
  \BibitemOpen
  \bibfield  {author} {\bibinfo {author} {\bibfnamefont {B.}~\bibnamefont
  {Swingle}},\ }\href {\doibase 10.1103/PhysRevD.86.065007} {\bibfield
  {journal} {\bibinfo  {journal} {Phys. Rev.}\ }\textbf {\bibinfo {volume}
  {D86}},\ \bibinfo {pages} {065007} (\bibinfo {year} {2012})},\ \Eprint
  {http://arxiv.org/abs/0905.1317} {arXiv:0905.1317 [cond-mat.str-el]}
  \BibitemShut {NoStop}%
%%CITATION = ARXIV:0905.1317;%%
\bibitem [{\citenamefont {Beny}(2013)}]{Beny:2011vh}%
  \BibitemOpen
  \bibfield  {author} {\bibinfo {author} {\bibfnamefont {C.}~\bibnamefont
  {Beny}},\ }\href {\doibase 10.1088/1367-2630/15/2/023020} {\bibfield
  {journal} {\bibinfo  {journal} {New J. Phys.}\ }\textbf {\bibinfo {volume}
  {15}},\ \bibinfo {pages} {023020} (\bibinfo {year} {2013})},\ \Eprint
  {http://arxiv.org/abs/1110.4872} {arXiv:1110.4872 [quant-ph]} \BibitemShut
  {NoStop}%
%%CITATION = ARXIV:1110.4872;%%
\bibitem [{\citenamefont {Sinai~Kunkolienkar}\ and\ \citenamefont
  {Banerjee}(2017)}]{SinaiKunkolienkar:2016lgg}%
  \BibitemOpen
  \bibfield  {author} {\bibinfo {author} {\bibfnamefont {R.}~\bibnamefont
  {Sinai~Kunkolienkar}}\ and\ \bibinfo {author} {\bibfnamefont
  {K.}~\bibnamefont {Banerjee}},\ }\href {\doibase 10.1142/S0218271817501437}
  {\bibfield  {journal} {\bibinfo  {journal} {Int. J. Mod. Phys.}\ }\textbf
  {\bibinfo {volume} {D26}},\ \bibinfo {pages} {1750143} (\bibinfo {year}
  {2017})},\ \Eprint {http://arxiv.org/abs/1611.08581} {arXiv:1611.08581
  [hep-th]} \BibitemShut {NoStop}%
%%CITATION = ARXIV:1611.08581;%%
\bibitem [{\citenamefont {Bao}\ \emph {et~al.}(2017{\natexlab{a}})\citenamefont
  {Bao}, \citenamefont {Cao}, \citenamefont {Carroll},\ and\ \citenamefont
  {Chatwin-Davies}}]{Bao:2017qmt}%
  \BibitemOpen
  \bibfield  {author} {\bibinfo {author} {\bibfnamefont {N.}~\bibnamefont
  {Bao}}, \bibinfo {author} {\bibfnamefont {C.}~\bibnamefont {Cao}}, \bibinfo
  {author} {\bibfnamefont {S.~M.}\ \bibnamefont {Carroll}}, \ and\ \bibinfo
  {author} {\bibfnamefont {A.}~\bibnamefont {Chatwin-Davies}},\ }\href
  {\doibase 10.1103/PhysRevD.96.123536} {\bibfield  {journal} {\bibinfo
  {journal} {Phys. Rev.}\ }\textbf {\bibinfo {volume} {D96}},\ \bibinfo {pages}
  {123536} (\bibinfo {year} {2017}{\natexlab{a}})},\ \Eprint
  {http://arxiv.org/abs/1709.03513} {arXiv:1709.03513 [hep-th]} \BibitemShut
  {NoStop}%
%%CITATION = ARXIV:1709.03513;%%
\bibitem [{\citenamefont {Bao}\ \emph {et~al.}(2017{\natexlab{b}})\citenamefont
  {Bao}, \citenamefont {Cao}, \citenamefont {Carroll},\ and\ \citenamefont
  {McAllister}}]{Bao:2017iye}%
  \BibitemOpen
  \bibfield  {author} {\bibinfo {author} {\bibfnamefont {N.}~\bibnamefont
  {Bao}}, \bibinfo {author} {\bibfnamefont {C.}~\bibnamefont {Cao}}, \bibinfo
  {author} {\bibfnamefont {S.~M.}\ \bibnamefont {Carroll}}, \ and\ \bibinfo
  {author} {\bibfnamefont {L.}~\bibnamefont {McAllister}},\ }\href@noop {} {\
  (\bibinfo {year} {2017}{\natexlab{b}})},\ \Eprint
  {http://arxiv.org/abs/1702.06959} {arXiv:1702.06959 [hep-th]} \BibitemShut
  {NoStop}%
%%CITATION = ARXIV:1702.06959;%%
\bibitem [{\citenamefont {Czech}\ \emph {et~al.}(2015)\citenamefont {Czech},
  \citenamefont {Lamprou}, \citenamefont {McCandlish},\ and\ \citenamefont
  {Sully}}]{Czech:2015qta}%
  \BibitemOpen
  \bibfield  {author} {\bibinfo {author} {\bibfnamefont {B.}~\bibnamefont
  {Czech}}, \bibinfo {author} {\bibfnamefont {L.}~\bibnamefont {Lamprou}},
  \bibinfo {author} {\bibfnamefont {S.}~\bibnamefont {McCandlish}}, \ and\
  \bibinfo {author} {\bibfnamefont {J.}~\bibnamefont {Sully}},\ }\href
  {\doibase 10.1007/JHEP10(2015)175} {\bibfield  {journal} {\bibinfo  {journal}
  {JHEP}\ }\textbf {\bibinfo {volume} {10}},\ \bibinfo {pages} {175} (\bibinfo
  {year} {2015})},\ \Eprint {http://arxiv.org/abs/1505.05515} {arXiv:1505.05515
  [hep-th]} \BibitemShut {NoStop}%
%%CITATION = ARXIV:1505.05515;%%
\bibitem [{\citenamefont {Czech}\ \emph
  {et~al.}(2016{\natexlab{a}})\citenamefont {Czech}, \citenamefont {Lamprou},
  \citenamefont {McCandlish},\ and\ \citenamefont {Sully}}]{Czech:2015kbp}%
  \BibitemOpen
  \bibfield  {author} {\bibinfo {author} {\bibfnamefont {B.}~\bibnamefont
  {Czech}}, \bibinfo {author} {\bibfnamefont {L.}~\bibnamefont {Lamprou}},
  \bibinfo {author} {\bibfnamefont {S.}~\bibnamefont {McCandlish}}, \ and\
  \bibinfo {author} {\bibfnamefont {J.}~\bibnamefont {Sully}},\ }\href
  {\doibase 10.1007/JHEP07(2016)100} {\bibfield  {journal} {\bibinfo  {journal}
  {JHEP}\ }\textbf {\bibinfo {volume} {07}},\ \bibinfo {pages} {100} (\bibinfo
  {year} {2016}{\natexlab{a}})},\ \Eprint {http://arxiv.org/abs/1512.01548}
  {arXiv:1512.01548 [hep-th]} \BibitemShut {NoStop}%
%%CITATION = ARXIV:1512.01548;%%
\bibitem [{\citenamefont {Czech}\ \emph
  {et~al.}(2016{\natexlab{b}})\citenamefont {Czech}, \citenamefont {Evenbly},
  \citenamefont {Lamprou}, \citenamefont {McCandlish}, \citenamefont {Qi},
  \citenamefont {Sully},\ and\ \citenamefont {Vidal}}]{Czech:2015xna}%
  \BibitemOpen
  \bibfield  {author} {\bibinfo {author} {\bibfnamefont {B.}~\bibnamefont
  {Czech}}, \bibinfo {author} {\bibfnamefont {G.}~\bibnamefont {Evenbly}},
  \bibinfo {author} {\bibfnamefont {L.}~\bibnamefont {Lamprou}}, \bibinfo
  {author} {\bibfnamefont {S.}~\bibnamefont {McCandlish}}, \bibinfo {author}
  {\bibfnamefont {X.-L.}\ \bibnamefont {Qi}}, \bibinfo {author} {\bibfnamefont
  {J.}~\bibnamefont {Sully}}, \ and\ \bibinfo {author} {\bibfnamefont
  {G.}~\bibnamefont {Vidal}},\ }\href {\doibase 10.1103/PhysRevB.94.085101}
  {\bibfield  {journal} {\bibinfo  {journal} {Phys. Rev.}\ }\textbf {\bibinfo
  {volume} {B94}},\ \bibinfo {pages} {085101} (\bibinfo {year}
  {2016}{\natexlab{b}})},\ \Eprint {http://arxiv.org/abs/1510.07637}
  {arXiv:1510.07637 [cond-mat.str-el]} \BibitemShut {NoStop}%
%%CITATION = ARXIV:1510.07637;%%
\bibitem [{\citenamefont {Preskill}(2018)}]{Preskill2018quantumcomputingin}%
  \BibitemOpen
  \bibfield  {author} {\bibinfo {author} {\bibfnamefont {J.}~\bibnamefont
  {Preskill}},\ }\href {\doibase 10.22331/q-2018-08-06-79} {\bibfield
  {journal} {\bibinfo  {journal} {{Quantum}}\ }\textbf {\bibinfo {volume}
  {2}},\ \bibinfo {pages} {79} (\bibinfo {year} {2018})}\BibitemShut {NoStop}%
\bibitem [{\citenamefont {Kogut}\ and\ \citenamefont
  {Susskind}(1975)}]{Kogut:1974ag}%
  \BibitemOpen
  \bibfield  {author} {\bibinfo {author} {\bibfnamefont {J.~B.}\ \bibnamefont
  {Kogut}}\ and\ \bibinfo {author} {\bibfnamefont {L.}~\bibnamefont
  {Susskind}},\ }\href {\doibase 10.1103/PhysRevD.11.395} {\bibfield  {journal}
  {\bibinfo  {journal} {Phys. Rev.}\ }\textbf {\bibinfo {volume} {D11}},\
  \bibinfo {pages} {395} (\bibinfo {year} {1975})}\BibitemShut {NoStop}%
%%CITATION = PHRVA,D11,395;%%
\bibitem [{\citenamefont {B.~Bravyi}\ and\ \citenamefont
  {Yu.~Kitaev}(2000)}]{Kitaev:2000ak}%
  \BibitemOpen
  \bibfield  {author} {\bibinfo {author} {\bibfnamefont {S.}~\bibnamefont
  {B.~Bravyi}}\ and\ \bibinfo {author} {\bibfnamefont {A.}~\bibnamefont
  {Yu.~Kitaev}},\ }\href {\doibase 10.1006/aphy.2002.6254} {\bibfield
  {journal} {\bibinfo  {journal} {Annals of Physics}\ }\textbf {\bibinfo
  {volume} {298}} (\bibinfo {year} {2000}),\
  10.1006/aphy.2002.6254}\BibitemShut {NoStop}%
\bibitem [{\citenamefont {Jordan}\ \emph {et~al.}(2011)\citenamefont {Jordan},
  \citenamefont {Lee},\ and\ \citenamefont {Preskill}}]{Jordan:2011ci}%
  \BibitemOpen
  \bibfield  {author} {\bibinfo {author} {\bibfnamefont {S.~P.}\ \bibnamefont
  {Jordan}}, \bibinfo {author} {\bibfnamefont {K.~S.~M.}\ \bibnamefont {Lee}},
  \ and\ \bibinfo {author} {\bibfnamefont {J.}~\bibnamefont {Preskill}},\
  }\href@noop {} {\  (\bibinfo {year} {2011})},\ \bibinfo {note} {[Quant. Inf.
  Comput.14,1014(2014)]},\ \Eprint {http://arxiv.org/abs/1112.4833}
  {arXiv:1112.4833 [hep-th]} \BibitemShut {NoStop}%
%%CITATION = ARXIV:1112.4833;%%
\bibitem [{\citenamefont {Martinez}\ \emph {et~al.}(2016)\citenamefont
  {Martinez} \emph {et~al.}}]{Martinez:2016yna}%
  \BibitemOpen
  \bibfield  {author} {\bibinfo {author} {\bibfnamefont {E.~A.}\ \bibnamefont
  {Martinez}} \emph {et~al.},\ }\href {\doibase 10.1038/nature18318} {\bibfield
   {journal} {\bibinfo  {journal} {Nature}\ }\textbf {\bibinfo {volume}
  {534}},\ \bibinfo {pages} {516} (\bibinfo {year} {2016})},\ \Eprint
  {http://arxiv.org/abs/1605.04570} {arXiv:1605.04570 [quant-ph]} \BibitemShut
  {NoStop}%
%%CITATION = ARXIV:1605.04570;%%
\bibitem [{\citenamefont {Muschik}\ \emph {et~al.}(2017)\citenamefont
  {Muschik}, \citenamefont {Heyl}, \citenamefont {Martinez}, \citenamefont
  {Monz}, \citenamefont {Schindler}, \citenamefont {Vogell}, \citenamefont
  {Dalmonte}, \citenamefont {Hauke}, \citenamefont {Blatt},\ and\ \citenamefont
  {Zoller}}]{Muschik:2016tws}%
  \BibitemOpen
  \bibfield  {author} {\bibinfo {author} {\bibfnamefont {C.}~\bibnamefont
  {Muschik}}, \bibinfo {author} {\bibfnamefont {M.}~\bibnamefont {Heyl}},
  \bibinfo {author} {\bibfnamefont {E.}~\bibnamefont {Martinez}}, \bibinfo
  {author} {\bibfnamefont {T.}~\bibnamefont {Monz}}, \bibinfo {author}
  {\bibfnamefont {P.}~\bibnamefont {Schindler}}, \bibinfo {author}
  {\bibfnamefont {B.}~\bibnamefont {Vogell}}, \bibinfo {author} {\bibfnamefont
  {M.}~\bibnamefont {Dalmonte}}, \bibinfo {author} {\bibfnamefont
  {P.}~\bibnamefont {Hauke}}, \bibinfo {author} {\bibfnamefont
  {R.}~\bibnamefont {Blatt}}, \ and\ \bibinfo {author} {\bibfnamefont
  {P.}~\bibnamefont {Zoller}},\ }\href {\doibase 10.1088/1367-2630/aa89ab}
  {\bibfield  {journal} {\bibinfo  {journal} {New J. Phys.}\ }\textbf {\bibinfo
  {volume} {19}},\ \bibinfo {pages} {103020} (\bibinfo {year} {2017})},\
  \Eprint {http://arxiv.org/abs/1612.08653} {arXiv:1612.08653 [quant-ph]}
  \BibitemShut {NoStop}%
%%CITATION = ARXIV:1612.08653;%%
\bibitem [{\citenamefont {Macridin}\ \emph
  {et~al.}(2018{\natexlab{a}})\citenamefont {Macridin}, \citenamefont
  {Spentzouris}, \citenamefont {Amundson},\ and\ \citenamefont
  {Harnik}}]{Macridin:2018gdw}%
  \BibitemOpen
  \bibfield  {author} {\bibinfo {author} {\bibfnamefont {A.}~\bibnamefont
  {Macridin}}, \bibinfo {author} {\bibfnamefont {P.}~\bibnamefont
  {Spentzouris}}, \bibinfo {author} {\bibfnamefont {J.}~\bibnamefont
  {Amundson}}, \ and\ \bibinfo {author} {\bibfnamefont {R.}~\bibnamefont
  {Harnik}},\ }\href {\doibase 10.1103/PhysRevLett.121.110504} {\bibfield
  {journal} {\bibinfo  {journal} {Phys. Rev. Lett.}\ }\textbf {\bibinfo
  {volume} {121}},\ \bibinfo {pages} {110504} (\bibinfo {year}
  {2018}{\natexlab{a}})},\ \Eprint {http://arxiv.org/abs/1802.07347}
  {arXiv:1802.07347 [quant-ph]} \BibitemShut {NoStop}%
%%CITATION = ARXIV:1802.07347;%%
\bibitem [{\citenamefont {Macridin}\ \emph
  {et~al.}(2018{\natexlab{b}})\citenamefont {Macridin}, \citenamefont
  {Spentzouris}, \citenamefont {Amundson},\ and\ \citenamefont
  {Harnik}}]{Macridin:2018oli}%
  \BibitemOpen
  \bibfield  {author} {\bibinfo {author} {\bibfnamefont {A.}~\bibnamefont
  {Macridin}}, \bibinfo {author} {\bibfnamefont {P.}~\bibnamefont
  {Spentzouris}}, \bibinfo {author} {\bibfnamefont {J.}~\bibnamefont
  {Amundson}}, \ and\ \bibinfo {author} {\bibfnamefont {R.}~\bibnamefont
  {Harnik}},\ }\href {\doibase 10.1103/PhysRevA.98.042312} {\bibfield
  {journal} {\bibinfo  {journal} {Phys. Rev.}\ }\textbf {\bibinfo {volume}
  {A98}},\ \bibinfo {pages} {042312} (\bibinfo {year} {2018}{\natexlab{b}})},\
  \Eprint {http://arxiv.org/abs/1805.09928} {arXiv:1805.09928 [quant-ph]}
  \BibitemShut {NoStop}%
%%CITATION = ARXIV:1805.09928;%%
\bibitem [{\citenamefont {Klco}\ \emph {et~al.}(2018)\citenamefont {Klco},
  \citenamefont {Dumitrescu}, \citenamefont {McCaskey}, \citenamefont {Morris},
  \citenamefont {Pooser}, \citenamefont {Sanz}, \citenamefont {Solano},
  \citenamefont {Lougovski},\ and\ \citenamefont {Savage}}]{Klco:2018kyo}%
  \BibitemOpen
  \bibfield  {author} {\bibinfo {author} {\bibfnamefont {N.}~\bibnamefont
  {Klco}}, \bibinfo {author} {\bibfnamefont {E.~F.}\ \bibnamefont
  {Dumitrescu}}, \bibinfo {author} {\bibfnamefont {A.~J.}\ \bibnamefont
  {McCaskey}}, \bibinfo {author} {\bibfnamefont {T.~D.}\ \bibnamefont
  {Morris}}, \bibinfo {author} {\bibfnamefont {R.~C.}\ \bibnamefont {Pooser}},
  \bibinfo {author} {\bibfnamefont {M.}~\bibnamefont {Sanz}}, \bibinfo {author}
  {\bibfnamefont {E.}~\bibnamefont {Solano}}, \bibinfo {author} {\bibfnamefont
  {P.}~\bibnamefont {Lougovski}}, \ and\ \bibinfo {author} {\bibfnamefont
  {M.~J.}\ \bibnamefont {Savage}},\ }\href {\doibase
  10.1103/PhysRevA.98.032331} {\bibfield  {journal} {\bibinfo  {journal} {Phys.
  Rev.}\ }\textbf {\bibinfo {volume} {A98}},\ \bibinfo {pages} {032331}
  (\bibinfo {year} {2018})},\ \Eprint {http://arxiv.org/abs/1803.03326}
  {arXiv:1803.03326 [quant-ph]} \BibitemShut {NoStop}%
%%CITATION = ARXIV:1803.03326;%%
\bibitem [{\citenamefont {Klco}\ and\ \citenamefont
  {Savage}(2018)}]{Klco:2018zqz}%
  \BibitemOpen
  \bibfield  {author} {\bibinfo {author} {\bibfnamefont {N.}~\bibnamefont
  {Klco}}\ and\ \bibinfo {author} {\bibfnamefont {M.~J.}\ \bibnamefont
  {Savage}},\ }\href@noop {} {\  (\bibinfo {year} {2018})},\ \Eprint
  {http://arxiv.org/abs/1808.10378} {arXiv:1808.10378 [quant-ph]} \BibitemShut
  {NoStop}%
%%CITATION = ARXIV:1808.10378;%%
\bibitem [{\citenamefont {Kaplan}\ and\ \citenamefont
  {Stryker}(2018)}]{Kaplan:2018vnj}%
  \BibitemOpen
  \bibfield  {author} {\bibinfo {author} {\bibfnamefont {D.~B.}\ \bibnamefont
  {Kaplan}}\ and\ \bibinfo {author} {\bibfnamefont {J.~R.}\ \bibnamefont
  {Stryker}},\ }\href@noop {} {\  (\bibinfo {year} {2018})},\ \Eprint
  {http://arxiv.org/abs/1806.08797} {arXiv:1806.08797 [hep-lat]} \BibitemShut
  {NoStop}%
%%CITATION = ARXIV:1806.08797;%%
\bibitem [{\citenamefont {Lieb}\ \emph {et~al.}(1961)\citenamefont {Lieb},
  \citenamefont {Schultz},\ and\ \citenamefont {Mattis}}]{Lieb:1961fr}%
  \BibitemOpen
  \bibfield  {author} {\bibinfo {author} {\bibfnamefont {E.~H.}\ \bibnamefont
  {Lieb}}, \bibinfo {author} {\bibfnamefont {T.}~\bibnamefont {Schultz}}, \
  and\ \bibinfo {author} {\bibfnamefont {D.}~\bibnamefont {Mattis}},\ }\href
  {\doibase 10.1016/0003-4916(61)90115-4} {\bibfield  {journal} {\bibinfo
  {journal} {Annals Phys.}\ }\textbf {\bibinfo {volume} {16}},\ \bibinfo
  {pages} {407} (\bibinfo {year} {1961})}\BibitemShut {NoStop}%
%%CITATION = APNYA,16,407;%%
\bibitem [{\citenamefont {Pfeuty}(1970)}]{Pfeuty:1970pf}%
  \BibitemOpen
  \bibfield  {author} {\bibinfo {author} {\bibfnamefont {P.}~\bibnamefont
  {Pfeuty}},\ }\href {\doibase 10.1016/0003-4916(70)90270-8} {\bibfield
  {journal} {\bibinfo  {journal} {Annals Phys.}\ }\textbf {\bibinfo {volume}
  {57}},\ \bibinfo {pages} {79} (\bibinfo {year} {1970})}\BibitemShut {NoStop}%
%%CITATION = APNYA,57,79;%%
\bibitem [{\citenamefont {Kitaev}(2006)}]{Kitaev:2006lla}%
  \BibitemOpen
  \bibfield  {author} {\bibinfo {author} {\bibfnamefont {A.}~\bibnamefont
  {Kitaev}},\ }\href {\doibase 10.1016/j.aop.2005.10.005} {\bibfield  {journal}
  {\bibinfo  {journal} {Annals Phys.}\ }\textbf {\bibinfo {volume} {321}},\
  \bibinfo {pages} {2} (\bibinfo {year} {2006})}\BibitemShut {NoStop}%
%%CITATION = APNYA,321,2;%%
\bibitem [{\citenamefont {Verstraete}\ \emph {et~al.}(2009)\citenamefont
  {Verstraete}, \citenamefont {Cirac},\ and\ \citenamefont
  {Latorre}}]{Verstraete:2008qpa}%
  \BibitemOpen
  \bibfield  {author} {\bibinfo {author} {\bibfnamefont {F.}~\bibnamefont
  {Verstraete}}, \bibinfo {author} {\bibfnamefont {J.~I.}\ \bibnamefont
  {Cirac}}, \ and\ \bibinfo {author} {\bibfnamefont {J.~I.}\ \bibnamefont
  {Latorre}},\ }\href {\doibase 10.1103/PhysRevA.79.032316} {\bibfield
  {journal} {\bibinfo  {journal} {Phys. Rev.}\ }\textbf {\bibinfo {volume}
  {A79}},\ \bibinfo {pages} {032316} (\bibinfo {year} {2009})},\ \Eprint
  {http://arxiv.org/abs/0804.1888} {arXiv:0804.1888 [quant-ph]} \BibitemShut
  {NoStop}%
%%CITATION = ARXIV:0804.1888;%%
\bibitem [{\citenamefont {Schmoll}\ and\ \citenamefont
  {Orus}(2017)}]{Schmoll:2016kbb}%
  \BibitemOpen
  \bibfield  {author} {\bibinfo {author} {\bibfnamefont {P.}~\bibnamefont
  {Schmoll}}\ and\ \bibinfo {author} {\bibfnamefont {R.}~\bibnamefont {Orus}},\
  }\href {\doibase 10.1103/PhysRevB.95.045112} {\bibfield  {journal} {\bibinfo
  {journal} {Phys. Rev.}\ }\textbf {\bibinfo {volume} {B95}},\ \bibinfo {pages}
  {045112} (\bibinfo {year} {2017})},\ \Eprint
  {http://arxiv.org/abs/1605.04315} {arXiv:1605.04315 [cond-mat.str-el]}
  \BibitemShut {NoStop}%
%%CITATION = ARXIV:1605.04315;%%
\bibitem [{\citenamefont {Jiang}\ \emph {et~al.}(2018)\citenamefont {Jiang},
  \citenamefont {Sung}, \citenamefont {Kechedzhi}, \citenamefont
  {Smelyanskiy},\ and\ \citenamefont {Boixo}}]{Jiang:2017pyp}%
  \BibitemOpen
  \bibfield  {author} {\bibinfo {author} {\bibfnamefont {Z.}~\bibnamefont
  {Jiang}}, \bibinfo {author} {\bibfnamefont {K.~J.}\ \bibnamefont {Sung}},
  \bibinfo {author} {\bibfnamefont {K.}~\bibnamefont {Kechedzhi}}, \bibinfo
  {author} {\bibfnamefont {V.~N.}\ \bibnamefont {Smelyanskiy}}, \ and\ \bibinfo
  {author} {\bibfnamefont {S.}~\bibnamefont {Boixo}},\ }\href {\doibase
  10.1103/PhysRevApplied.9.044036} {\bibfield  {journal} {\bibinfo  {journal}
  {Phys. Rev. Applied}\ }\textbf {\bibinfo {volume} {9}},\ \bibinfo {pages}
  {044036} (\bibinfo {year} {2018})},\ \Eprint
  {http://arxiv.org/abs/1711.05395} {arXiv:1711.05395 [quant-ph]} \BibitemShut
  {NoStop}%
%%CITATION = ARXIV:1711.05395;%%
\end{thebibliography}
\end{document}